\documentclass[aps,prl,floatfix,reprint,showpacs,superscriptaddress,longbibliography]{revtex4-2}
\usepackage{mathtools,amssymb,graphicx,units}
\usepackage[plainpages=false,pdfpagelabels,colorlinks=true,linkcolor=red,urlcolor=blue,citecolor=blue,pdftitle={Title},pdfauthor={},pdfdisplaydoctitle=true,pdfduplex=DuplexFlipLongEdge]{hyperref}
\usepackage{soul} 
\usepackage[dvipsnames,usenames]{color}
\usepackage[normalem]{ulem}
\usepackage{graphicx}
\usepackage{xcolor}
\usepackage{bbold, verbatim} 
\usepackage{float}
\usepackage[font=small,justification=justified]{subcaption} 
\emergencystretch=\maxdimen
\usepackage[colorinlistoftodos,prependcaption,textsize=tiny]{todonotes}
\usepackage{cprotect}
\usepackage{ulem}
\usepackage{siunitx}
\usepackage{xr}
\def\s{\sigma}

\begin{document}
\title{High-pressure hydrogen by machine learning  
and quantum Monte Carlo}

\author{Andrea Tirelli} 
\email{atirelli@sissa.it}
\affiliation{International School for Advanced Studies (SISSA),
Via Bonomea 265, 34136 Trieste, Italy}
\author{Giacomo Tenti} 
\affiliation{International School for Advanced Studies (SISSA),
Via Bonomea 265, 34136 Trieste, Italy}
\author{Kousuke Nakano} 
\affiliation{International School for Advanced Studies (SISSA),
Via Bonomea 265, 34136 Trieste, Italy}
\affiliation{School of Information Science, JAIST, Asahidai 1-1, Nomi, Ishikawa 923-1292, Japan}
\author{Sandro Sorella} 
\email{sorella@sissa.it}
\affiliation{International School for Advanced Studies (SISSA),
Via Bonomea 265, 34136 Trieste, Italy}
\affiliation{Computational Materials Science Research Team, RIKEN Center for Computational Science (R-CCS), Kobe, Hyogo 650-0047, Japan}

\begin{abstract} 
We have developed a technique combining the accuracy of quantum Monte Carlo in describing the electron correlation with the
efficiency of a Machine Learning Potential (MLP). We use  kernel regression in combination with  SOAP (Smooth Overlap of Atomic Position) features, implemented here in a very efficient way. 
The key ingredients are: 
i) a sparsification technique, based on farthest point sampling, 
ensuring  generality and transferability of our MLPs 
and ii) the so called $\Delta$-learning, allowing 
a  small training data  set, a 
fundamental property for 
highly  accurate but computationally demanding calculations, such as the ones based on quantum Monte Carlo. As the first application we present a benchmark study of the liquid-liquid transition of high-pressure hydrogen and show the quality of our MLP, by emphasizing the importance of high accuracy for  this very debated subject, where experiments are
difficult in the lab, and theory is still far from being conclusive.
\end{abstract}
\date{\today}
\maketitle
\textbf{Introduction.}
 %
The determination of the phase diagram of hydrogen at high pressure has been a particularly long-standing problem in the physics community since the pioneering work of Wigner and Huntington~{\cite{1935WIG}}. 
Recently it has  received renewed interest,
\textit{e.g.}, for the still controversial properties  of the newly discovered H-based superconductors such as H$_{3}$S~{\cite{2015DRO}} and LaH$_{10}$~{\cite{2019SOM}}
, for establishing the polymorphs of the solid phase of hydrogen 
{\cite{drummond2015,2019DRO}} and, last but not least, for characterizing 
  the nature of the liquid-liquid transition (LLT) between molecular and atomic phases. 
  
  In this work we are mainly addressing the last aspect, 
  that is of paramount importance  for determining the composition of giant planets, and, by consequence, of our universe.
 Unfortunately, many contradictory                             experimental~{\cite{2018CEL, 2015KNU, 2015OHT, 2016MCW, 2016ZAG, 2017ZAG}} and computational~{\cite{2003SCA, 2010MOR, 2010LOR, 2006DEL, 2007VOR, 2018MAZ, 2019GEN, 2020CHE}} results have been reported so far, about whether the LLT is a      first-order or a smooth transition, and the location of the LLT on the $p$-$T$ phase diagram. 
Indeed, for studying     phase transition phenomena,  molecular dynamics simulations  with long times and  several atoms are   required, in order to exclude  equilibration and finite-size effect artifacts, respectively.
 In this respect, %
 the construction of Machine Learning Potentials (MLPs), aimed to 
 reproduce  accurate electronic energies and force fields,  is one of the most promising approaches to overcome the large computational cost of present {\it ab-initio} techniques for electronic simulation~{\cite{2020JIA}}. In fact, very recently, Cheng et al.~{\cite{2020CHE}} constructed MLPs for hydrogen, using the Behler--Parrinello artificial Neural Network framework (NNPs)~{\cite{2007BEH, 2011BEH}} based on Density Functional Theory (DFT) references. 
 Here we show that  this approach can be extended to a state-of-the-art 
 wave function method, quantum Monte Carlo (QMC)~{\cite{2001FOU}}, that  is well known to describe very accurately the electronic correlation but has a computational cost 
 at  least three orders of magnitude larger than DFT.
 Indeed in hydrogen, it is important to have a very reliable technique, 
 because experiments are difficult at extreme conditions and DFT, on the other hand, depends too much on the functional used~{\cite{2013SAM, 2014RAY}}.
 In this letter we aim to produce a MLP
 that is efficient enough to allow long and large size simulations, but on the other hand, retains the same accuracy of QMC, and, importantly, the training can be achieved by a relatively small number of simulations. 
 %
 %
 We will show that this is possible, by employing  the $\Delta$-learning method combined with the  Gaussian Kernel Regression (GKR) framework~{\cite{2010BAR}}.  \\

\textbf{Model and methods.}
\label{sec:model}
We outline the theoretical and technical details of the Machine Learning (ML) models that can be trained using our algorithmic approach.\\ 
\textit{SOAP kernel method.}
One of the key features of modern MLPs is their applicability to systems with an arbitrary number of atoms $N$, allowing simulations with sizes much larger than the ones used during the training. One possibility is to write the total energy of the system as a sum of contributions depending on the local environment around each atom placed at  the position $\mathbf{r}_i$ \cite{2007BEH}, $  E = \sum_{i = 1}^N e(R_i) $ where $R_i = \left(\mathbf{r}_1  -\mathbf{r}_i  , \dots ,\mathbf{r}_N -\mathbf{r}_i  \right) \equiv \left( \mathbf{r}_{1i} , \dots, \mathbf{r}_{Ni}\right)$. The function $e(R_i)$ depends only on the neighbors of atom $i$ inside a sphere with radius $r_c$, namely such that $r_{ji}=|\mathbf{r}_{ji}|\le r_c$ ($j=1,\cdots, N$).
Several approaches are possible to parametrize $e(R_i)$. In this work we used a kernel regression model 
\begin{align}
    e(R_i) = \sum_{j = 1}^{N_{\rm{train}}} c_{j} \mathcal{K} \left(R_i ,R_j \right) \label{local energy kernel},
\end{align}
where $\mathcal{K}\left( R_i , R_j\right)$ is the normalized kernel between two local environments $R_i$ (the target one) and $R_j$ (the ones used for the training, that may or may not contain $R_i$), and $c_{j}$ are coefficients to be determined. Here $N_{\rm{train}}$ is the  total number of local environments used for the construction of the MLP.
%
%

The local environment around each atom is characterized using the Smooth Overlap of Atomic Positions (SOAP), \cite{2013BAR}: in such an approach, a density function $\rho(R_i,\mathbf{r}) $ is associated to each local environment $R_i$. Hence:
\begin{align*}
    \rho(R_i,\mathbf{r}) = \sum\limits_{j| r_{ij}\le r_c}
    f_c(r_{ij}) \exp\left( -\frac{( \mathbf{r}_{ij} - \mathbf{r})^2}{2\s^2}\right),
\end{align*}
where $f_c$ is a cutoff function and $\sigma$ is a hyper-parameter of the model. In this work a piece-wise defined polynomial, vanishing smoothly at $r_{ij} = r_c$, was used as $f_c(r_{ij})$ \footnote{$f(r_{ij})=1$ if $0<r_{ij}\le r_c/2$ and $f(r_{ij})=4 (r_{ij}/r_c-1)^2 (4 r_{ij}/r_c -1)$ for $r_c/2\le r_{ij}\le r_c$. We also assume that $f(0)=0$, in order to exclude the trivial $i=j$ contribution.}.  The \emph{similarity kernel} between two local environments $R_i$ and $R_j$ is defined as the average over all rotations $\hat{U}$ of the overlap between the densities raised to some exponent $n$:
\begin{align}
    \Tilde{k}(R_i,R_j) = \int d \hat{U} \left( \int d \mathbf{r}  \rho(R_i,\mathbf{r}) \rho(\hat{U} R_j,\mathbf{r}) \right)^{n}. \label{SOAP similarity kernel}
\end{align}
With this definition the resulting kernel is explicitly invariant under permutation of the particles and rotations in 3D space. In order to compute the integral in Eq.~\eqref{SOAP similarity kernel}, the usual approach is to expand $\rho$ in radial and angular components and evaluate  the power spectrum \cite{2013BAR}. 
Here instead, in analogy with the nowadays  standard implementation  of non local pseudopotentials within QMC \cite{mitas_pseudo} (where a similar angular
integration appears),
we  have used  an approximate version of Eq.~\eqref{SOAP similarity kernel} which takes the average of the overlap only over a finite
subgroup (e.g. all the symmetries of the cube) of $SO(3)$, that is spanned by $n_{\rm{sym}}$  $3\times 3$ real orthogonal matrices 
$U_k$:
\begin{align} \label{eq:sumk}
    \Tilde{k}(R_i,R_j) = \frac{1}{n_{\rm{sym}}} \sum_{k=1}^{n_{\rm{sym}}}
    \left( \int d \mathbf{r}  \rho(R_i,\mathbf{r}) \rho(\hat{U}_k R_j,\mathbf{r}) \right)^{n}.
\end{align}
The resulting kernel is explicitly invariant under the set of transformations considered, that  approaches the full continuous symmetry  
for large $n_{\rm{sym}}$.
Just like for QMC pseudopotential calculations, we have experienced that a relatively small $n_{\rm{sym}}$ is sufficient to reach the desired accuracy.  
Moreover Eq. (\ref{eq:sumk}) can be computed in a very efficient way since each term in the sum has a simple expression in terms of Gaussian functions. 
Finally, in order to obtain $\mathcal{K}$ in Eq.~\eqref{local energy kernel}, we normalize $\Tilde{k}$ and raise it to some power $\eta$, so that $ 0 \leq \mathcal{K} \leq 1$. \\

\textit{Local sparsification.}\label{subsec:sparse}
A key aspect in the implementation and training of a ML model is the prevention of overfitting, \textit{i.e.} the statistical phenomenon for which a model can fit almost exactly the training data but fails to predict unknown data reliably. 

Overfitting problems can arise when the  number of variational parameters of a given model is too large. Moreover, since the number of variational parameters in a GKR model depends linearly on the size of the training set, particular care should be taken when selecting the corresponding reference samples. This becomes even more important in the present setting: indeed, adding one configuration with $N$ atoms to the training dataset implicitly defines 
$N$ new local environments corresponding to further $N$
variational degrees of freedom of the model in Eq.~\eqref{local energy kernel}. Additionally, the computational effort needed in the training of a GKR model is polynomial  (at most cubic) in the size of the training set -- while instead stochastic optimization algorithms employed in neural networks allow the complexity to be independent of the size of the number of samples.  

Therefore, within GKR, the careful choice of the  local environments  used in the training phase is of paramount importance. This is usually obtained by means of the so  called  \textit{sparsification}. We implement a variant of it, that we name \textit{local sparsification}, since we discard from each configuration some or even a large fraction of the corresponding 
$N$ local environments. This  prevents overfitting and allows an increased predictive power of the model on unseen data.

The local sparsification preprocessing can be performed with multiple algorithms, and we employ the \textit{farthest point sampling scheme}, which is a greedy algorithm first adopted as an approximate solution of the so-called Center Selection Problem (see \cite[\S 11.2]{kleinberg2006algorithm}). 
A detailed explanation of our algorithm is contained in the Supplementary Material. \\

\begin{figure*}[t]
	\centering
	\begin{subfigure}{.48\textwidth}
		\centering
		\includegraphics[width=\linewidth]{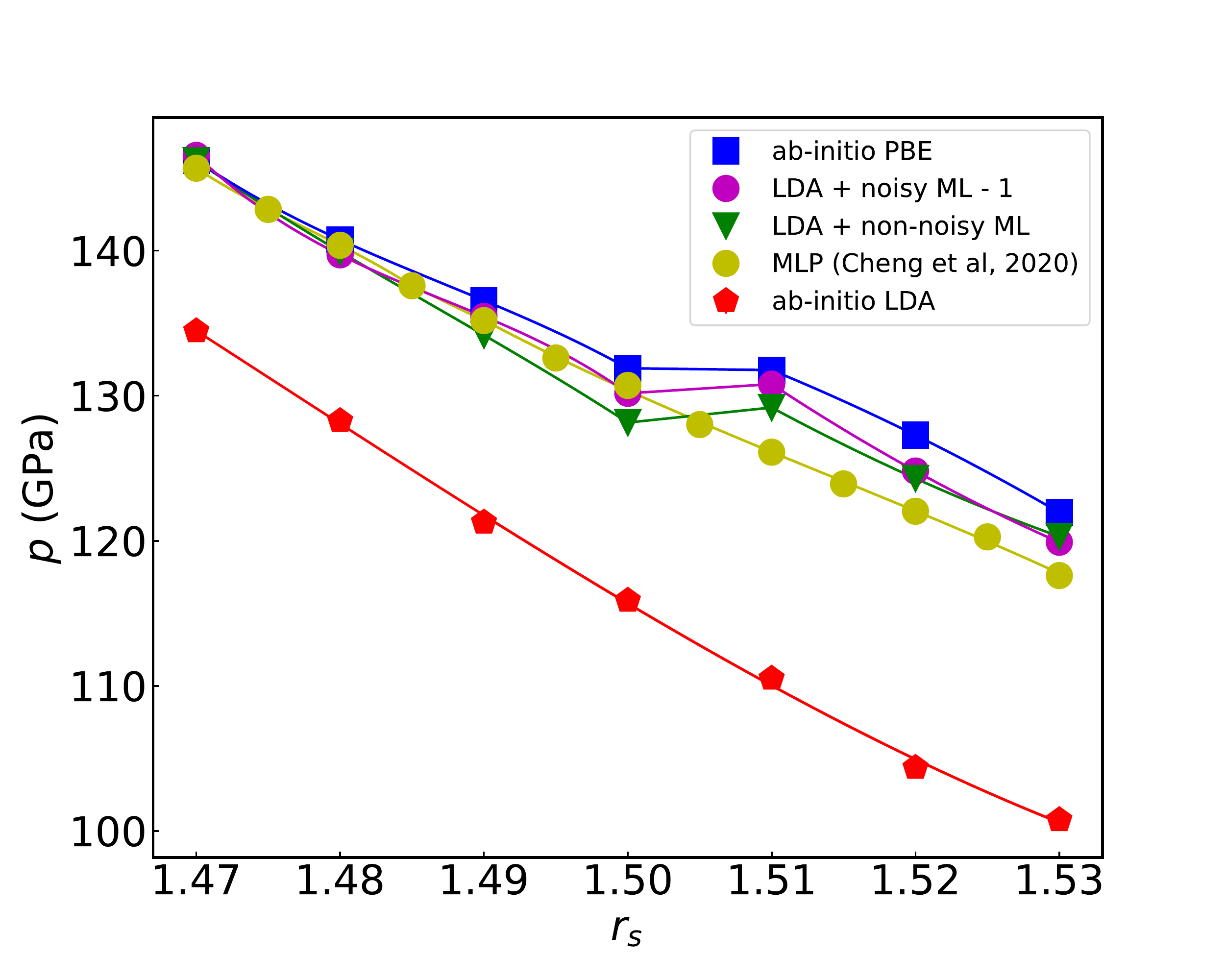}
		\caption{}
		\label{fig:llt_pbe}
	\end{subfigure}
	\begin{subfigure}{.48\textwidth}
		\centering
		\includegraphics[width=\linewidth]{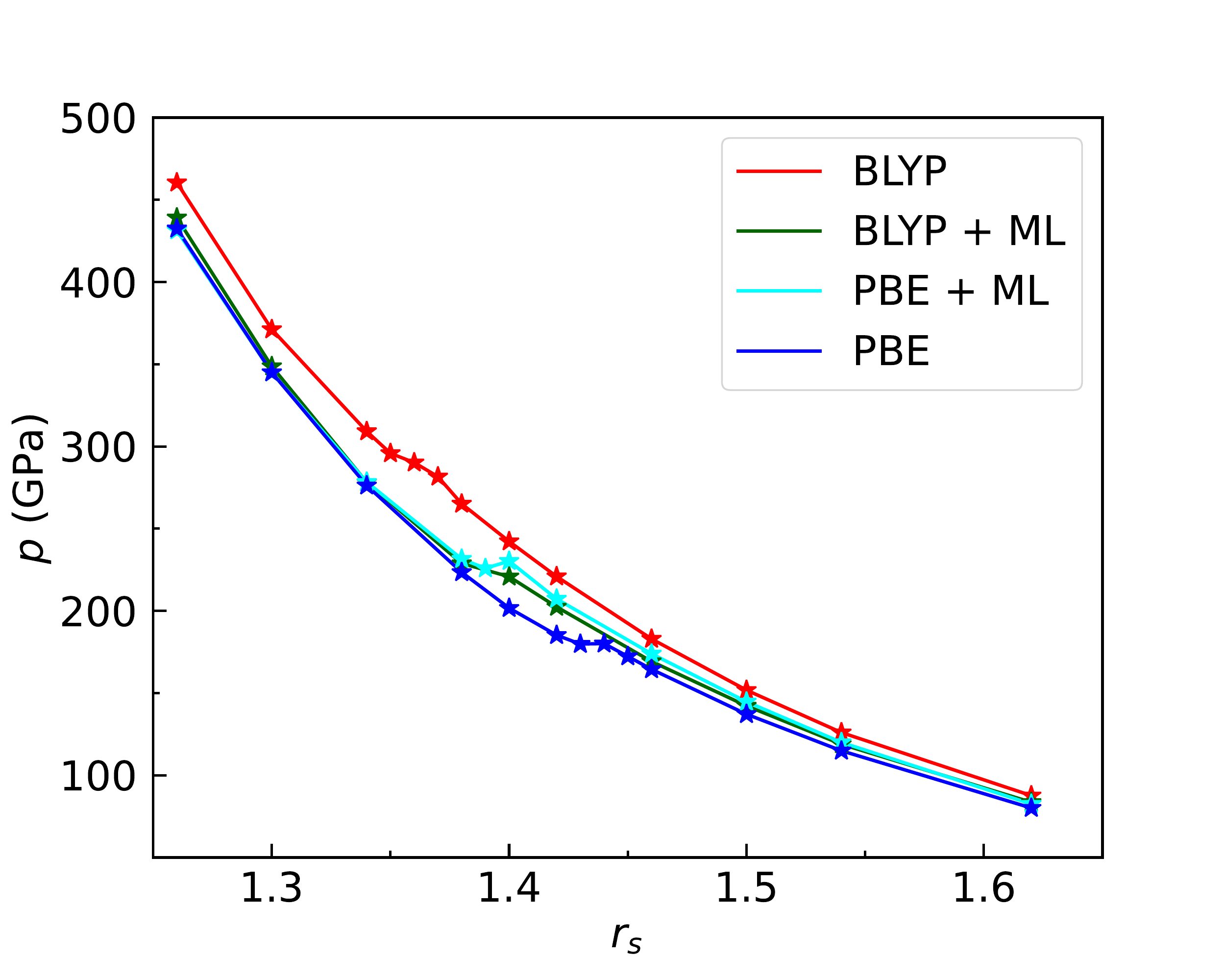}
		\caption{}
		\label{fig:comparison}
	\end{subfigure}
	\caption{ (a) Comparison between molecular dynamics  simulations via MLPs (cyan and green lines for our GKR, yellow line for the NN-MLP of Chen et al.~\cite{2020CHE} and the AIMD PBE and LDA references (blue and red lines, respectively) for 128 H at 1400~K. The values of the hyper-parameters used for all the MLPs are reported in the SI. (b) EOS of 128 H at 1000K using DFT with PBE and BLYP functionals and the appropriate ML corrections.}

	\label{fig:comp}
\end{figure*}
\textit{$\Delta$-machine learning.}
In this work we also follow the $\Delta$-Machine Learning ($\Delta$-ML) approach, first introduced in Ref.~\onlinecite{2015RAM} as a composite strategy
reproducing a target \textit{ab-initio} method, by adding
ML corrections to a much cheaper baseline reference potential. Here, we take as baseline \textit{ab-initio} reference  a DFT calculation with  PBE functional and as target method Variational Monte Carlo (VMC). 

There is a significant gain in exploiting the $\Delta$-ML approach: by training on the difference between two \textit{ab-initio} methods, the machine learning task becomes easier. This is crucial for the case at hand: if we were to train a GKR model to predict VMC observables, the amount of training configurations to achieve  the desired accuracy would require a computational effort which is at present not feasible. In particular, we are able to obtain a VMC potential with an accuracy comparable with the error bar of the VMC predictions, by training the model on a relatively small dataset. Moreover, the computational cost of the ML inference (also thanks to the local sparsification preprocessing) is negligible compared to the baseline DFT calculation. Hence,  we are able to compute observables with VMC accuracy at the cost of a DFT simulation. \\

\textbf{Results.}
\label{sec:zerot}
We discuss now  the first successful  application of our method to the study of the LLT in hydrogen at high pressures. Our dataset is composed of $684$ configurations of $128$ hydrogen atoms in  cubic simulation cells of different edge sizes, from $10.56$ to $11.53$ bohr. 
For each of these configurations we computed the ground state energy, the force components acting on each atom and the pressure of the system using two different \textit{ab-initio} methods: DFT with the PBE and BLYP functionals, with the \verb|Quantum Espresso| software package, \cite{2009GIA}; VMC, with the \verb|TurboRVB| software package, \cite{2020NAK2}. Concerning the VMC simulations, the predicted energies, force components and pressure have average error bars of $2\times 10^{-3}$ Ha,  $1.8\times 10^{-3}$ Ha/Bohr and  $0.1$ GPa, respectively. Note that performing a molecular dynamics (MD) simulation using VMC with such an accuracy at each step would be computationally prohibitive.\\ 


\begin{figure*}[th]
    \centering
    \begin{subfigure}{.48\textwidth}
  \centering
  \includegraphics[width=\linewidth]{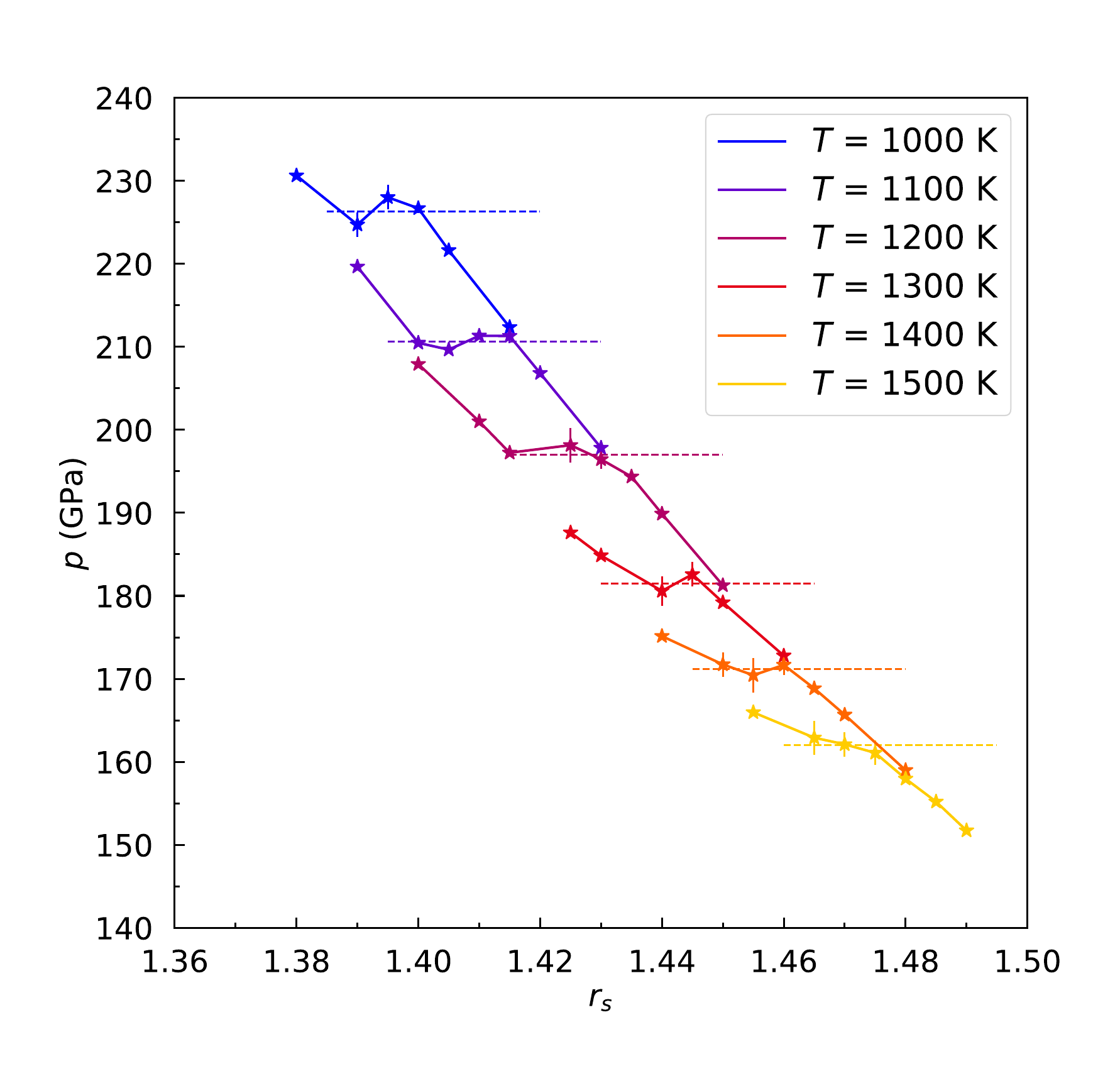}
  \caption{}
  \label{fig:isotherms_256}
\end{subfigure}
    \begin{subfigure}{.48\textwidth}
  \centering
  \includegraphics[width=\linewidth]{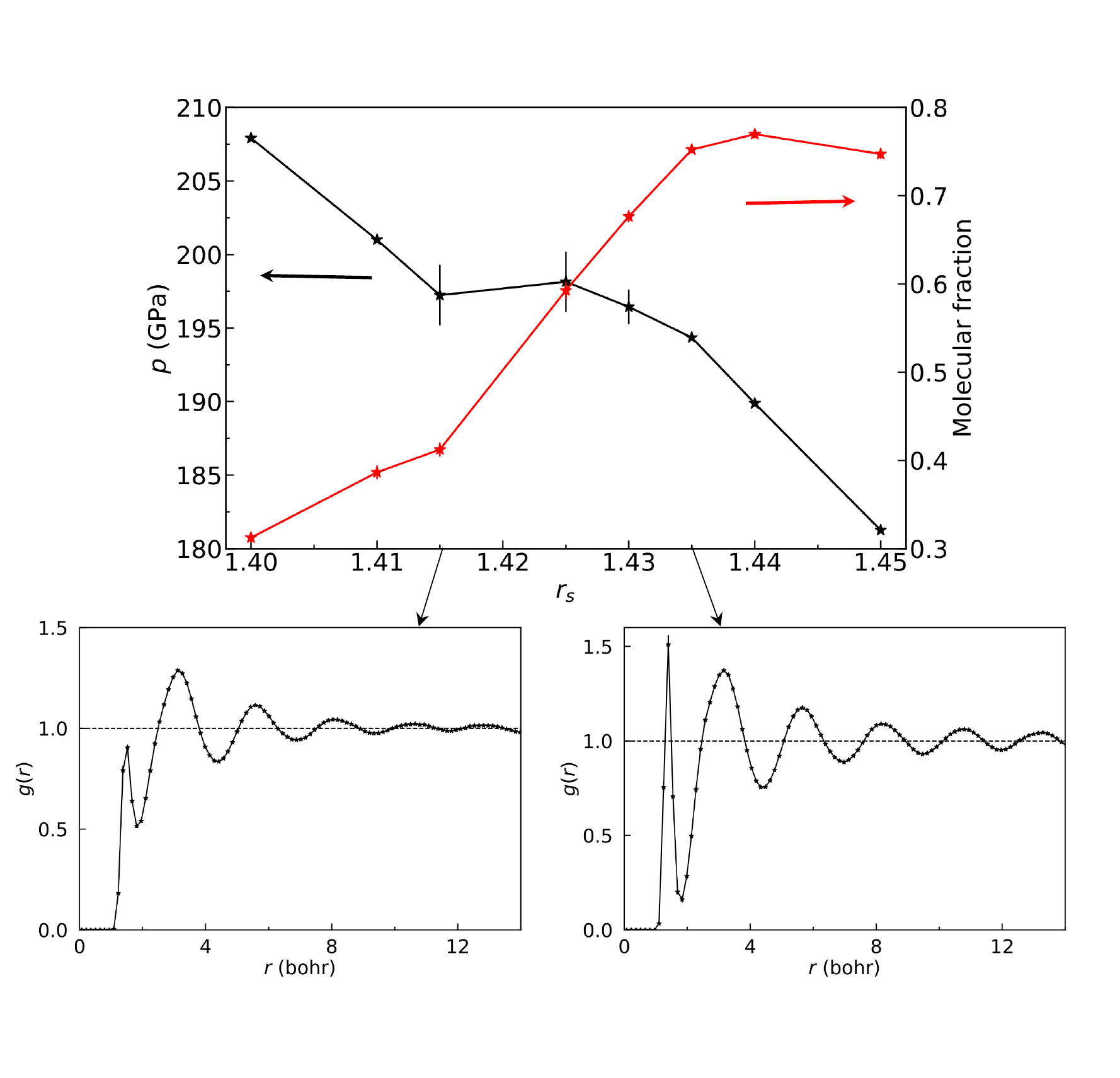}
  \caption{}
  \label{fig:isotherm1200_256_with_gr}
\end{subfigure}
\caption{(a) Isotherm curves at different values of temperature for a system of 256 hydrogen atoms. Dashed lines represent the transition pressure obtained when $p$ becomes constant (within the error bars) in some range of densities. (b) Isotherm for $T$ = 1200~K (black line). The molecular fraction (defined in the text) as a function of the Wigner Seitz radius $r_s$ (red line) and the radial distribution functions for $r_s$ = 1.415 and $r_s$ = 1.435 (boxes in the bottom) are also shown.}
\label{fig:isotherm_results}
\end{figure*}

\textit{Performances of the $\Delta$-ML correction.}
Although computing the root mean square error (RMSE) on target quantities is a standard approach to access the predictive power of a ML model, it seems necessary in our case to have more tailored qualitative and quantitative performance metrics.
Since QMC provides training data with intrinsic noise, the first step in our study is to determine how this affects the ML approach.
To this purpose, 
since it is not possible to have an unbiased --i.e. noiseless--  VMC reference, we have performed the following test: 
 we trained our GKR on two datasets, such that one was simply the difference between a DFT baseline reference (calculated with the LDA functional) and a DFT-PBE target, while the other was a noisy version of the first, where we introduced Gaussian white noises on the observables in order to emulate   QMC  inputs.
As shown in Fig. \ref{fig:llt_pbe} the effect of a noise $\simeq 10^{-3}$ a.u. on forces and total energies \footnote{This error bar can be achieved for instance by a single QMC simulation on about one thousand processors for about a couple of hours within the TurboRVB package \cite{2020NAK2}. Clearly this error bar cannot be sustained for an MD run with several thousands steps, as the ones employed here.}
is almost negligible, yielding a pressure plateau almost indistinguishable  from the reference PBE one and the MLP trained with the noiseless data set. 
The performances of the proposed MLP are also much better than the ones of the previous attempt \cite{2020CHE}, unable to reproduce the pressure plateau in this case. This is remarkable because in this region 
the  LDA baseline is very far from the reference  and also featureless. Note also that our finding is compatible with Fig. S6 of Ref.~\onlinecite{2020CHE}, since in that graph the $r_{\rm{s}}$ grid employed for the Ab-initio Molecular Dynamics (AIMD) PBE simulations is very coarse and away from the plateau.

Given this remarkable  stability property of our  GKR against noise, we further show the independence of the method (see Fig.~\ref{fig:comparison})  when trained 1) with true QMC noisy data 
 and 2) starting from two references, namely PBE and BLYP.
As shown in Fig.~\ref{fig:comparison},  
we have obtained an excellent  agreement between the two MLP EOS, even  though  the two DFT-based isotherms differ significantly from each other. In particular the transition pressure changes by  almost $100$ GPa with  the two chosen DFT functionals. Nevertheless the two models corrected with the MLP are much closer, with a difference in the predicted $p$ not larger than $3-4$ GPa (or anyway within the error bars) in the range of densities explored. 
For the rest of this work, we used the PBE-DFT method as our baseline potential, since it appears to provide results closer to the VMC ones. \\

\textit{LLT for high-pressure hydrogen.}
We have considered two systems of 128 and 256 hydrogen atoms respectively and performed a series of classical NVT simulations for several values of the density and temperature. 
At each step of MD we first compute DFT-PBE forces for the given configuration and then sum the correction given by our MLP.
We have used a time step of 10 a.u. (i.e. 0.242 fs) and controlled the temperature using either second order Langevin dynamics \cite{VANKAMPEN2007219} or a Bussi-Parrinello thermostat \cite{bussi-parrinello}.
A cubic supercell has been considered for all the simulations. 
For each point we have chosen a time simulation length between 2~ps and 10~ps, from which we have calculated the average of the thermodynamic quantities of interest. 
We remark  that, even though our $\Delta$-ML approach is still expensive (it has the same cost of a standard DFT Born-Oppenheimer MD), it is about $10^3$ times faster than computing directly VMC forces at each step, with the chosen accuracy. 

For all cases considered  we have found a narrow but clear plateau,  suggesting a first order nature of the LLT transition in this system (Fig.~\ref{fig:isotherms_256}). Both the transition pressures and corresponding densities are quantitatively  much different from the DFT ones, suggesting the importance of the VMC correction for the accurate evaluation of those quantities. 

We have also computed the average of the molecular fraction along the trajectory, defined as the percentage of atoms with at least one neighbor inside a cutoff of $1.4$ a.u., i.e. 
\begin{align*}
    m = \frac{1}{N} \sum_i g_c\left( \min_{j, j \neq i} | \mathbf{r}_i - \mathbf{r}_j |\right),
\end{align*}
where $g_c(r)$ is a cutoff function that is equal to $1$ for $r < 1.4$ a.u. and goes smoothly to $0$ for $r > 1.6$ a.u. 

As evident in Fig. (\ref{fig:isotherm1200_256_with_gr}), at the location of the pressure plateau the molecular fraction jumps from small values, corresponding to an atomic liquid, to values closer to $1$, typical for a molecular system. 
This can be also noticed by computing the radial distribution function for two values of the density at opposite sides of the transition.

\begin{figure}[t]
    \centering
    \includegraphics[width = 0.48\textwidth]{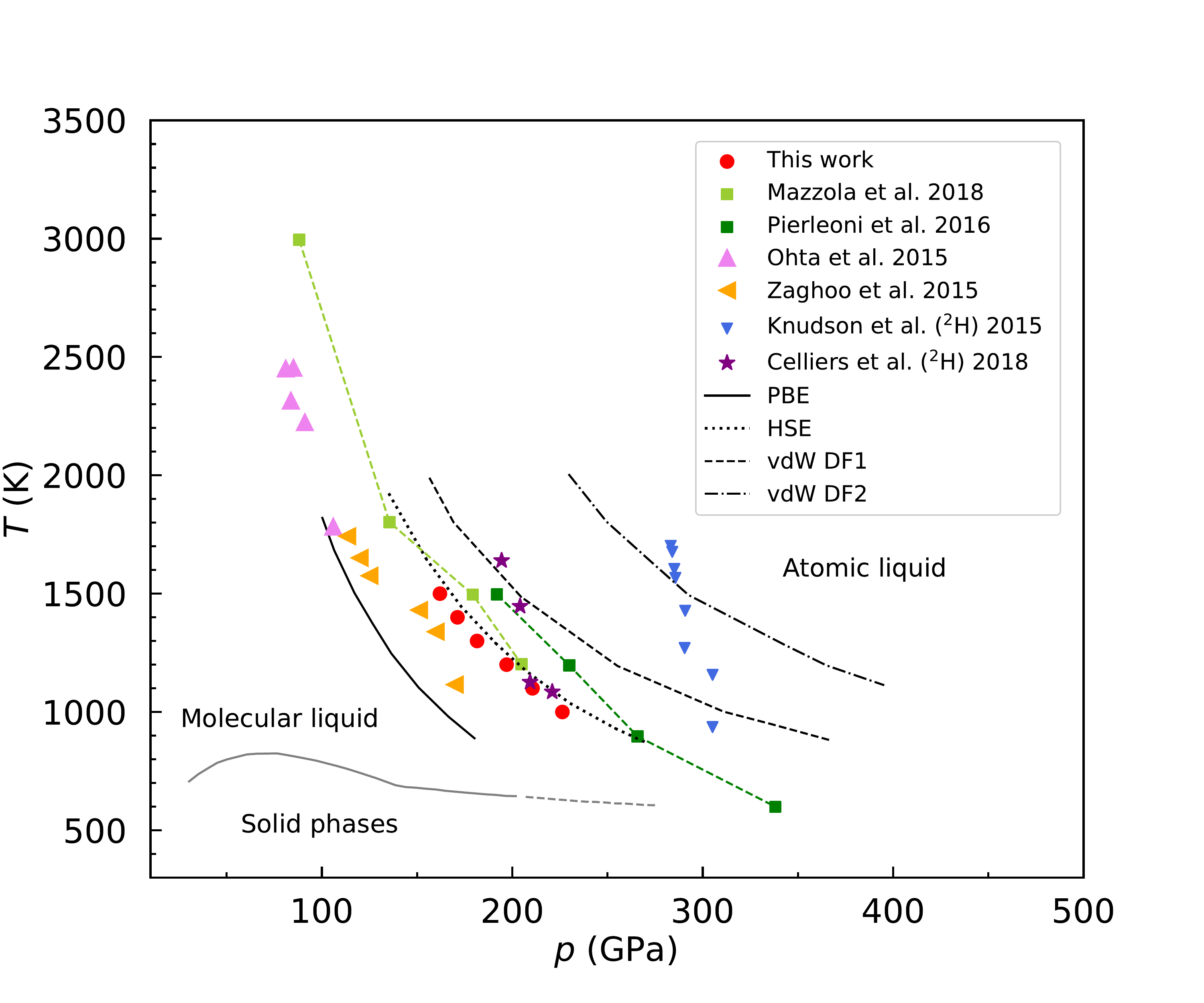}
    \caption{Proposed transition lines of the LLT in hydrogen. Our results  (red circles) . Previous (classical) QMC calculations taken from Refs.~\onlinecite{2018MAZ} and \onlinecite{2016PIE} (light green and dark green squares). DFT results for several functionals taken from Ref.~\onlinecite{2015KNU} (black lines). DAC experimental data taken from Refs.~\onlinecite{2015OHT} and \onlinecite{2016ZAG} (pink and orange triangles). Dynamical compression experiments taken from Refs.~\onlinecite{2015KNU} and \onlinecite{2018CEL} (blue triangles and violet stars, respectively). The literature data were digitalized using {\textsc{WebPlotDigitizer~\cite{2020ROH}.}}}
    \label{fig:llt_transition}
\end{figure}

In Fig. (\ref{fig:llt_transition}) we report the computed transition line, 
alongside with previous numerical and experimental results. As expected, we have found good agreement with previous VMC based molecular dynamics simulations that used the same wave function \cite{2018MAZ} but the statistical error on energy, forces and pressure were an order of magnitude larger, thus explaining the small differences found. 
Our calculations give transition pressures that are ~ 30 GPa below the ones predicted by the CEIMC method \cite{2016PIE}. The reason of this discrepancy is not clear at present; in principle the calculations done in Ref.~\onlinecite{2016PIE} could be more accurate due to the presence of backflow correlation and Diffusion Monte Carlo (DMC) projection, that could be further included for the training 
of our MLP. However, from our benchmark tests DMC gives only a small correction to the VMC forces, that is  compatible with the error bars as shown in the Supplementary Material. 
Our results are also in excellent agreement with the Diamond Anvil Cell (DAC) experiments of Refs.~\onlinecite{2015OHT} and \onlinecite{2016ZAG}, especially if we take into account that quantum corrections tend to further lower the transition pressure. 

We conclude this discussion by pointing out that our method gives almost the same line as the hybrid HSE functional. This could be useful for future  studies based on  DFT for  high pressure hydrogen.  \\

\textbf{Conclusion. } 
In  this work we have introduced  a successful approach, combining the ML and QMC techniques: in particular, through the $\Delta$-ML framework, we are able to obtain energies, forces and pressure predictions with QMC accuracy, at a computational cost comparable to  DFT, allowing simulations that were previously impossible (see SI Fig.~S7, where MD simulations with $N>1000$, at the QMC accuracy, are shown). Our technique is perfectly suited for being employed in a context where high accuracy is needed, but small training data sets are available.
We benchmarked our approach by studying the LLT in hydrogen at high pressures, whose phase diagram has been under a very lively debate, especially recently. In this context, our method is capable   
of reproducing the  pressure plateau computed directly with PBE at 1400K, in contrast to  the previously proposed NN MLP. This outcome appears in agreement with Ref.~\onlinecite{karasiev2020} but the subtle question whether a genuine first order transition or a sharp crossover occurs, is outside the  scope of this letter, and will be dealt with in future works with much larger $N$ and simulation times. Moreover 
we believe that further substantial improvements will be possible by using 
baseline potentials (e.g. the recent ultrafast MLP \cite{xie2021ultrafast}) much more efficient than DFT ones. 
Indeed, 
we expect that our approach will enable the study of systems that have been so far impossible to simulate, due to the almost intractable computational cost required by advanced many-body techniques, such as QMC. \\

\textbf{Acknowledgments.}
We thank Matthias Rupp for useful discussions and careful reading of the manuscript. 
The computations in this work have been mainly performed using the supercomputer Fugaku provided by RIKEN through the HPCI System Research Project (Project ID: hp210038) and Marconi100 provided by CINECA through the PRACE project No.~{2019204934} and ISCRA project AI-H-QMC, HP10BGJH1X.
K.N. is also grateful for computational resources from the facilities of Research Center for Advanced Computing Infrastructure at Japan Advanced Institute of Science and Technology (JAIST).
A.T. and S.S. acknowledge  financial  support  from the  MIUR  Progetti  di  Ricerca  di  Rilevante  Interesse Nazionale  (PRIN)  Bando  2017  -  grant  2017BZPKSZ.
K.N. acknowledges a support from the JSPS Overseas Research Fellowships, that from Grant-in-Aid for Early-Career Scientists Grant Number JP21K17752, and that from Grant-in-Aid for Scientific Research(C) Grant Number JP21K03400.
This work is supported by the European Centre of Excellence in Exascale Computing TREX - Targeting Real Chemical Accuracy at the Exascale. This project has received funding from the European Union’s Horizon 2020 - Research and Innovation program - under grant agreement no. 952165.

\bibliography{references}

\end{document}


\title{Supplemental material: High-pressure hydrogen by machine learning  
	and quantum Monte Carlo}
\author{Andrea Tirelli} 
\email{atirelli@sissa.it}
\affiliation{International School for Advanced Studies (SISSA),
Via Bonomea 265, 34136 Trieste, Italy}
\author{Giacomo Tenti} 
\affiliation{International School for Advanced Studies (SISSA),
Via Bonomea 265, 34136 Trieste, Italy}
\author{Kousuke Nakano} 
\affiliation{International School for Advanced Studies (SISSA),
Via Bonomea 265, 34136 Trieste, Italy}
\affiliation{School of Information Science, JAIST, Asahidai 1-1, Nomi, Ishikawa 923-1292, Japan}
\author{Sandro Sorella} 
\email{sorella@sissa.it}
\affiliation{International School for Advanced Studies (SISSA),
Via Bonomea 265, 34136 Trieste, Italy}
\affiliation{Computational Materials Science Research Team, RIKEN Center for Computational Science (R-CCS), Kobe, Hyogo 650-0047, Japan}

\date{\today}

\maketitle

\makeatletter
\renewcommand{\refname}{}
\renewcommand*{\citenumfont}[1]{S#1}
\renewcommand*{\bibnumfmt}[1]{[S#1]}
\makeatother

\setcounter{table}{0}
\setcounter{equation}{0}
\setcounter{figure}{0}
\renewcommand{\thetable}{S\Roman{table}}
\renewcommand{\thefigure}{S\arabic{figure}}
\renewcommand{\theequation}{S\arabic{equation}}
This supplemental material provides the theoretical and technical details on a number of aspects mentioned in the main paper: we provide information on the construction of the training datasets; we carefully explain how we performed the training of the MLPs; we give some technical insights on the implementation and parallelization of our algorithms. Lastly, we provide some results for the validation tests conducted to assess the quality of out MLPs.

\section{Dataset construction}
\subsection{Consistency of Variational and Diffusion Monte Carlo}
In this section we present several results regarding the reliability of the forces obtained with VMC. 
VMC forces, unlike DFT ones, do not depend on the particular functional for the calculation, but they depend only on the quality of the wavefunction. 

By differentiating the expression of the variational energy, one obtains
\begin{align}
    \mathbf{F}^{\rm{VMC}}_i = - \frac{\partial E^{\rm{VMC}}}{d\mathbf{r}_i} = \mathbf{F}^{\rm{HF}}_i + \mathbf{F}^{\rm{P}}_i -\sum_j \frac{\partial E^{\rm{VMC}}}{\partial \alpha_j } \frac{d\alpha_j}{d\mathbf{r}_i}\label{vmcforce_express}
\end{align}
where $\mathbf{F}^{\rm{HF}}_i$ and $\mathbf{F}^{\rm{P}}_i$ are the Hellman-Feynman and the Pulay terms respectively~{\cite{2020NAK2, 2021SWCT}}, and ${\mathbf{r}}$ refers to an atomic coordinate. 
The last term in Eq.~\eqref{vmcforce_express} becomes zero when the WF is optimized with respect to all the variational parameters.
Our choice of using the Jastrow-Slater WF can, in principle, introduce a small bias in the force, the so-called \emph{self consistency error}, since the determinant part is not optimized. In order to check if this is the case, we have calculated one component of the forces by fitting the potential energy surface (PES) of the system as a function of the displacement of one of the atoms and compared this value with one obtained directly from VMC force evaluation. We have repeated this procedure for two configurations at different densities ($r_s = 1.90$ and $r_s = 1.26$) and for different choices of the basis set used in the determinant part. For these calculations we have used twisted boundary condition with two different values of the twist $k$ (in particular $k = \frac{2\pi}{a}(\frac{1}{4},\frac{1}{4},\frac{1}{4})$ for $r_s = 1.90$ and $k = \frac{2\pi}{a}(\frac{1}{8},\frac{1}{8},\frac{1}{8})$ for $r_s= 1.26$). The results are shown in Tab.~{\ref{consistency_result}}. 
As we can clearly see, for both configurations the force is consistent with the value estimated with the PES for all the bases. Moreover the value computed directly does not depend on the chosen basis (with an accuracy of $\sim 10^{-3}$  Ha/Bohr), thus corroborating our choice to use a fairly small basis composed of three $s$-orbitals. 

\begin{table}[H]
    \centering
    \begin{tabular}{c|ccc|c|ccc}
    \multicolumn{1}{c}{}&  \multicolumn{3}{c}{\textbf{QMC}} & \multicolumn{4}{c}{\textbf{DFT}} \vspace{4mm}\\
    & VMC & VMC-PES & DMC-PES & LDA-PES (TurboRVB) & LDA (Q-E) & PBE (Q-E) & BLYP (Q-E)\\
    \hline
    3$s$ & 0.0543(7) & 0.055(1) & - & 0.05372 &\multirow{4}{*}{0.053610}& \multirow{4}{*}{0.053334}& \multirow{4}{*}{0.054877}\\
    3$s$1$p$ & 0.0531(9) & 0.051(1) & - & 0.05341 & \\
    4$s$2$p$ & 0.0542(4) & 0.052(1) & 0.0518(14) & 0.05321  \\
    4$s$2$p$1$d$ & 0.0538(4) & 0.051(1) & - & 0.05331\\
    \end{tabular}
    \cprotect\caption{Computed forces acting on a Hydrogen atom in H$_{128}$ with $r_s$ = 1.26, and the corresponding values obtained by fitting the PESs. 
     PBE and BLYP forces were calculated using \verb|Quantum Espresso| (\verb|QE|). LDA was computed both by \verb|QE| and \verb|TurboRVB| packages (the latter one in  the corresponding atomic basis).}
    \label{consistency_result}
\end{table}
\begin{table}[H]
    \centering
    \begin{tabular}{c|ccc|c|ccc}
    \multicolumn{1}{c}{}&  \multicolumn{3}{c}{\textbf{QMC}} & \multicolumn{4}{c}{\textbf{DFT}} \vspace{4mm}\\
    & VMC & VMC-PES & DMC-PES & LDA-PES (TurboRVB) & LDA (Q-E) & PBE (Q-E) & BLYP (Q-E)\\
    \hline
    3$s$   & -0.1595(7) & -0.1625(34) & - & -0.1667 &\multirow{4}{*}{-0.16604}& \multirow{4}{*}{-0.15859}& \multirow{4}{*}{-0.15356}\\
    3$s$1$p$ & -0.1603(6) & -0.1607(34) & - & -0.1667 & \\
    4$s$2$p$ & -0.1612(6) & -0.1638(33) & -0.1601(14) & -0.1665 \\
    4$s$2$p$1$d$ & -0.1614(7) & -0.1645(32) & - & -0.1663 \\
    \end{tabular}
    \cprotect\caption{Computed forces acting on a Hydrogen atom in H$_{128}$ with $r_s$ = 1.90, and the corresponding values obtained by fitting the PESs. PBE and BLYP forces were calculated using \verb|QE|. LDA was computed both by \verb|QE| and \verb|TurboRVB| packages (the latter one in  the corresponding atomic basis).}
    \label{results}
\end{table}

\begin{figure*}[t]
    \centering
    \begin{subfigure}{.48\textwidth}
  \centering
  \includegraphics[width=\linewidth]{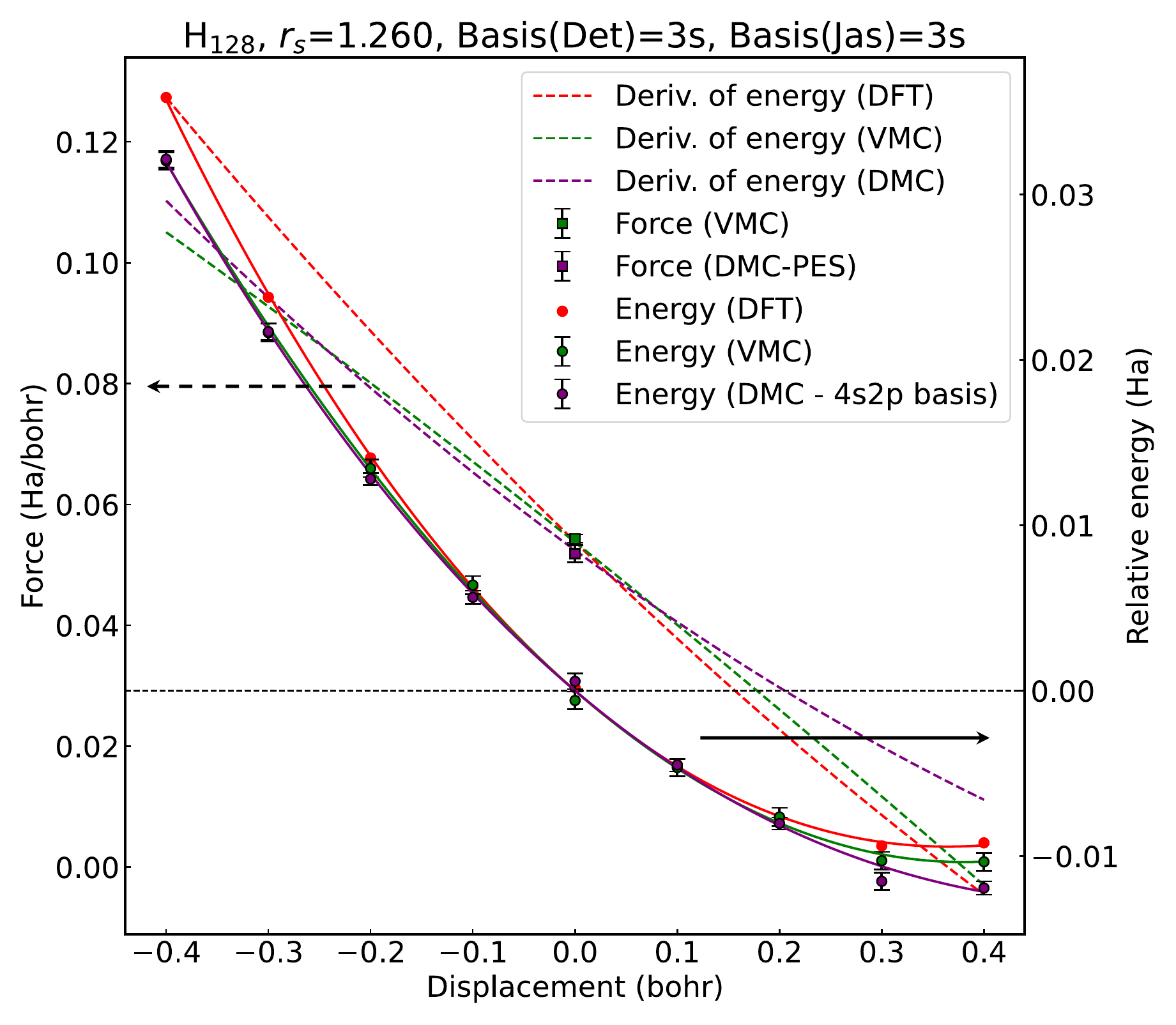}
  \caption{}
  \label{fig:consistency_results_rs_126}
\end{subfigure}
    \begin{subfigure}{.48\textwidth}
  \centering
  \includegraphics[width=\linewidth]{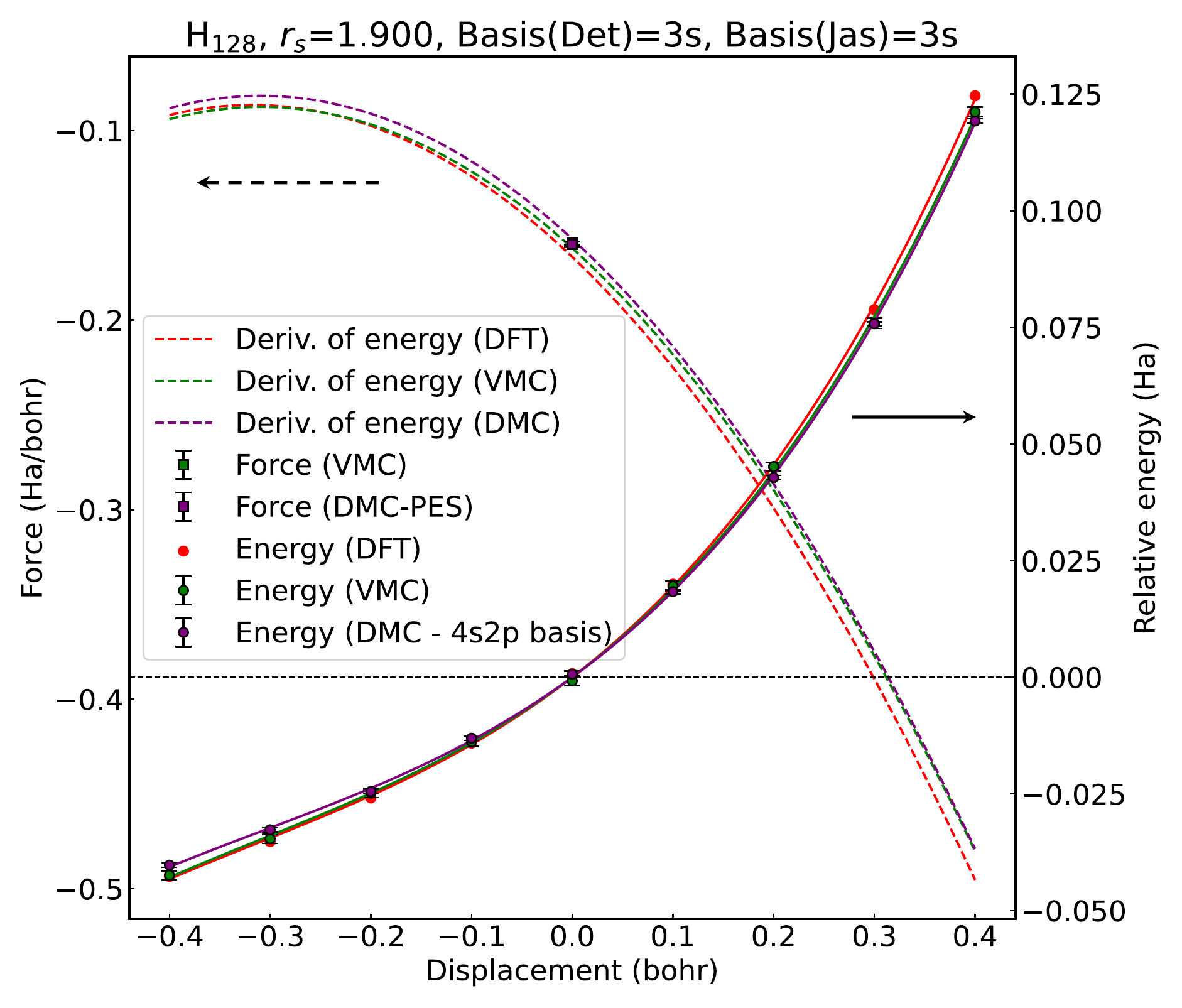} 
  \caption{}
  \label{fig:consistency_results_rs_190}
\end{subfigure}
\caption{Potential energies obtained by DFT, VMC and LRDMC calculations, their derivatives and the corresponding forces. The DFT and VMC calculations were done with 3$s$ basis, while LRDMC calculations were done with 4$s$2$p$ basis. (a) H$_{128}$ with $r_s$ = 1.26 and (b) H$_{128}$ with $r_s$ = 1.90.}
\label{fig:consistency_results}
\end{figure*}

To study the consistency of our VMC forces with the exact value, we have also computed the diffusion Monte Carlo (DMC) PES, using as guiding WF the JDFT solution obtained with the $(4s2p)$ basis and applying the lattice regularized diffusion Monte Carlo (LRDMC)~\cite{2020NAK2} method. The results (see Tab.~{\ref{consistency_result}}) show good agreement within the statistical errors for the $r_s = 1.90$ configuration and very small discrepancy ($\sim 2 \times 10^{-3}$ Ha/bohr) in the $r_s = 1.26$ case. The detailed comparisons in the 3$s$ cases are shown in Fig.~{\ref{fig:consistency_results}}. From this we can deduce that, in high pressure hydrogen, DMC is expected to give only a small correction with respect to the VMC force. Notice that this property is not trivial, and in general DMC and VMC forces can be very different.

\subsection{Training set construction}
We have ’initialized’ our training set by selecting $510$ liquid configurations with $128$ atoms from the dataset used by Ref.~\onlinecite{2020CHE}, namely  containing configurations from PBE-MD simulations at T ranging from 800K to 1500K, and with $r_s \in  [1.26, 1.6]$. In particular, we selected configurations such that the $r_s$ distribution was similar to the one of the original dataset.

For each configuration we have calculated DFT-PBE energies, forces and pressures using \verb|QE|, with a plane wave cut-off of $60$ Ry. To obtain the VMC forces, we have optimized a Jastrow-Slater wave function using $500$ steps with the stochastic reconfiguration method and then calculated forces and  pressures with \verb|TurboRVB|. As discussed previously, we have taken a basis composed of three $s$ orbitals for both the determinant and the Jastrow part. As good practice to minimize size effects \cite{2016PIE} we have also used twisted average boundary conditions (TABC) over a $4 \times 4 \times 4 $ Monkhorst-Pack mesh of the Brillouin zone for both DFT and QMC calculations.

During the MD simulations, other configurations were added to the dataset when needed, following an iterative procedure. At the end, we have used a dataset containing $684$ configurations to construct the final model. In order to check whether the prediction of the MLP on a given configuration could be trusted or not, we have used an \emph{extrapolation warning} criterion based on the similarity kernel. In particular we have defined the quantity
\begin{equation*}
\chi = \frac{1}{N} \sum_{i=1}^N \max_{j} K(R_i,R^{\mathrm{train}}_j) 
\end{equation*}
where $R_1,\dots,R_N$ and $R^{\mathrm{train}}_1 , \dots , R^{\mathrm{train}}_{N_{\mathrm{train}}}$ are the local environments of the configuration and the training set respectively. Notice that, if most of the $R_i$ are new to the MLP, $\chi$ will be a small number. In practice we added a new configuration to the training set when the measured value of $\chi$ was less than $ 0.75$. Although empirical, this criterion has been proven to be effective.

\section{Theory and implementation of Gaussian Kernel Regression}
\subsection{Sparsification}
The logic behind the Furthest Point Selection (FPS) method is the following: given a point cloud $X$, in the first iteration, select a random point $x \in X$ and add it to the point selection $S$; from the second iteration onward, iteratively choose the point from $X\setminus S$ which is as far as possible from the set $S$ at the present iteration, \textit{i.e.} we select a point $y\in X\setminus S$ such that 
\begin{equation}\label{eq:fps}
y = \underset{{y \in X\setminus S}}{\mathrm{argmax}}\ d(y, S),
\end{equation}
where $d$ is a distance on $X$, and, by definition, $d(y, S) = \inf_{x\in S}d(x, y)$.

In our context, the point cloud $X$ is given by the set of all local environments in the training set: namely, letting the indices $i$ and $j$ range over configurations and local environments respectively, we have that $X=\{ l_{i, j} \}_{i =1, \dots, N, j=1,\dots, N_i}$, where $N$ is the total number of configurations and $N_i$ is the number of atoms in configuration $i$. The key point is to understand when two such local environments are \textit{far} from each other: this is achieved with the kernel defined by the equation
\begin{align}\label{norm-kernel}
    \mathcal{K} (R_i ,R_j) = \left( \frac{\Tilde{k} ( R_i, R_j) }{\sqrt{\Tilde{k}(R_i, R_i) \Tilde{k}(R_j , R_j)}}\right)^{\eta},
\end{align} which gives a similarity score between two local environments, with value in the unit interval. Since similarities measures are dual to distance functions, we need to modify the original criterion of FPS given in Eq.~\eqref{eq:fps}: instead of choosing the point $y$ maximizing the distance from the current set $S$, one has to choose the point $y$ which minimizes the similarity measure $\mathcal{K}$ from $S$; this is possible since $\mathcal{K}$ induces a similarity score $\Bar{\mathcal{K}}$ between points and set, defined as 
\[
\Bar{\mathcal{K}}(y, S)= \sup_{x\in S} \mathcal{K}(x, y). 
\]
Therefore, Eq.~\eqref{eq:fps} becomes
\begin{equation}
y = \underset{{y \in X\setminus S}}{\mathrm{argmin}}\ \Bar{\mathcal{K}}(y, S).
\end{equation}
Thus, the FPS algorithm iteratively enriches the set $S$ by choosing local environments which make $S$ as diverse as possible. Local environments are added until a stopping criterion $\mathcal{C}$ is met. In our implementation of FPS, two possible choices of $\mathcal{C}$ are available: 
\begin{itemize}
    \item the number of local environments in $S$ reaches a user defined value
    \item the value $\min_{y\in X\setminus S} \Bar{\mathcal{K}}(y, S)$ is greater or equal than a user defined threshold; in this case, the number of local environments in $S$ is automatically set by the algorithm. 
\end{itemize}
The latter choice is in general preferable, as the number of selected local environments strongly depends on the diversity of the dataset: indeed, given a fixed similarity threshold, if the dataset is composed by very correlated configurations, then the number of retained local environments will be much smaller than for non-correlated configurations. Typical values for the similarity threshold are in a range between 0.75 and 0.9.

Note that, in order to run the FPS algorithm, regardless of what stopping criterion is adopted, the matrix $M$ of pairwise similarities between local environments has to be computed: namely, $M$ is given by entries $m_{i, j}$, where
\[
m_{i, j} = \mathcal{K}(l_i, l_j),
\]
where both $i$ and $j$ run from $1$ to $N_l$, where $N_l=\sum_{i=1}^N N_i$ is the total number of local environments. Since $N_l$ can be a very large number (even for small number of configurations, if they contain a large number of atoms), FPS has been implemented in a distributed way, so that, in a parallel execution of the algorithm, the allocation of the matrix $M$ is distributed over the number of processes. This is necessary in order to be able to work with large size datasets. 

An important feature of our implementation of FPS to the present setting is that the selection of relevant data points is performed at the level of local environments: from one configuration, we are able to select specific local environments so that the sparsification procedure is as fine and precise as possible. 
\begin{figure*}[t]
	\centering
	\begin{subfigure}{.48\textwidth}
		\centering
		\includegraphics[width=\linewidth]{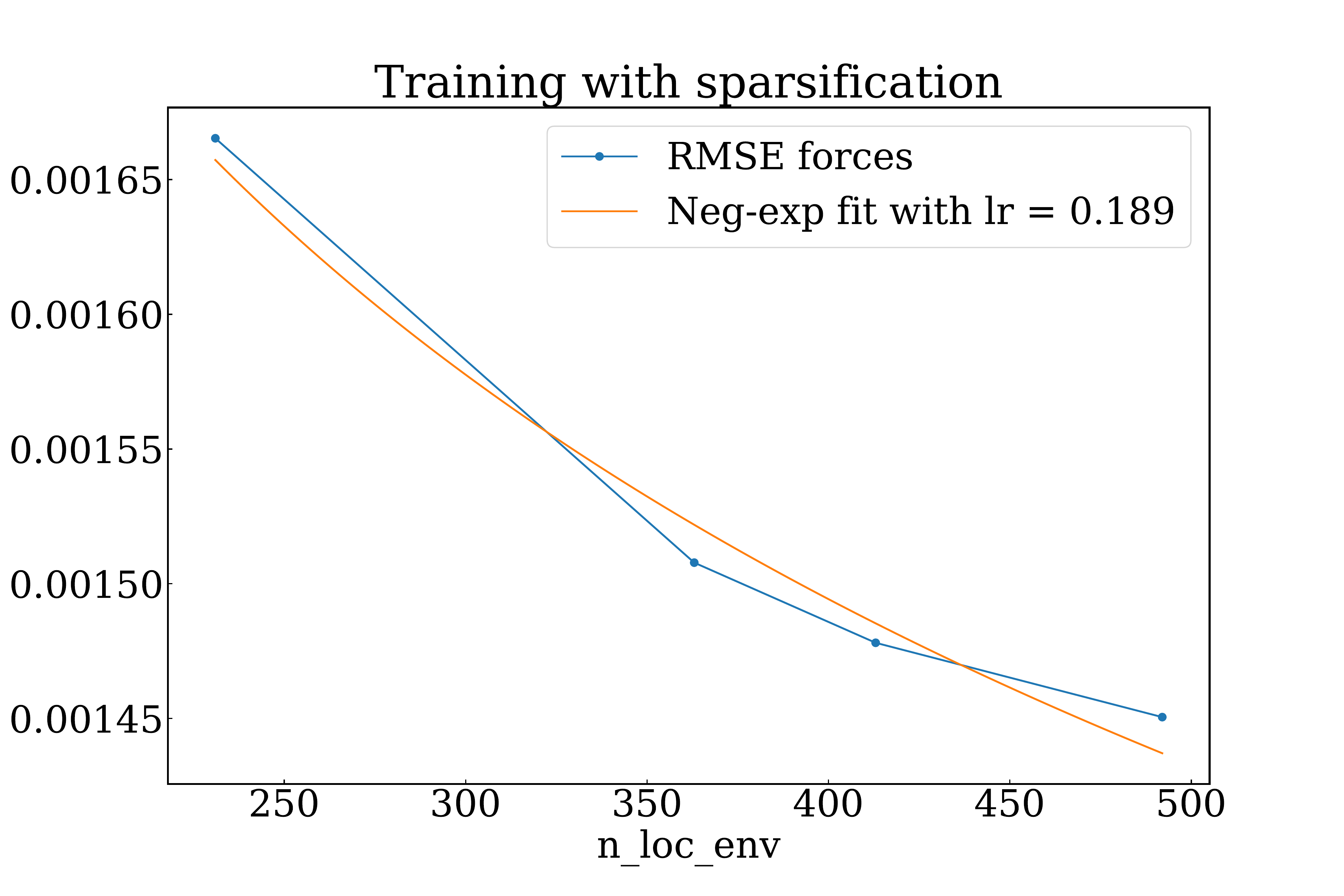}
		\caption{}
		\label{fig:lr_sparse}
	\end{subfigure}
	\begin{subfigure}{.48\textwidth}
		\centering
		\includegraphics[width=\linewidth]{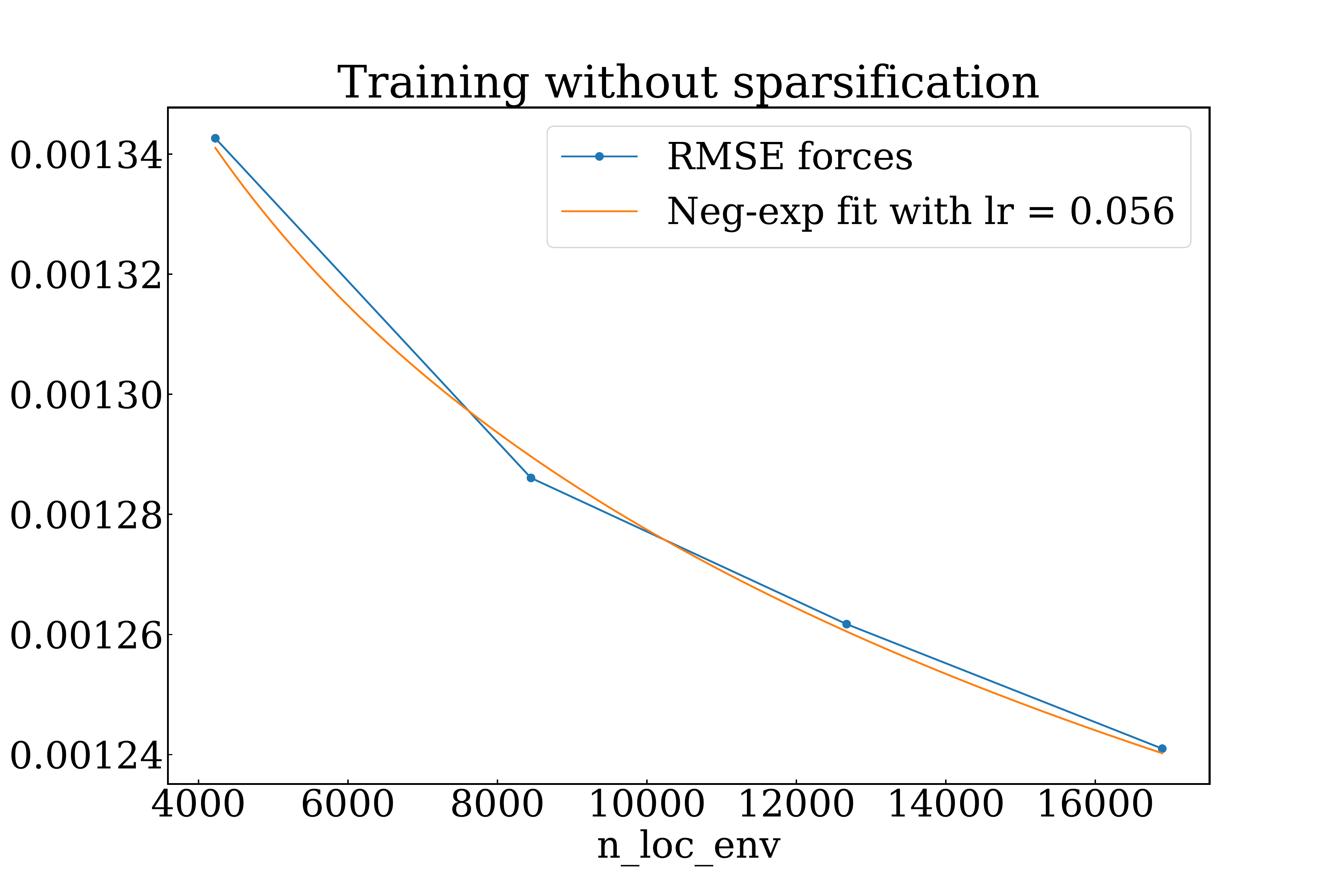} 
		\caption{}
		\label{fig:lr_no_sparse}
	\end{subfigure}
	\caption{RMSE on forces as a function of the number of local environments. (a) Training using the sparsification preprocessing step. (b) Training without using the sparsification preprocessing step. In both cases, the blue line represents the observed values, whereas the orange line is the interpolation with a negative exponential function, the exponent of which is normally called {\it learning rate} (lr). The reported RMSEs are evaluated on the test set, none of whose configurations has been seen by the model during training. }
	\label{fig:lr_sparsification}
\end{figure*}
We performed some tests to access the robustness of the sparsification procedure in terms of model performances and the gain in terms of computational resources used for the training of the model on the sparsified dataset. Fig.~{\ref{fig:lr_sparsification}} shows that the sparsification procedure does not have significant effects on the RMSE but allows training on the same dataset in a much smaller amount of time: while performances differ only by $16 \%$, the number of local environments needed without sparsification is about 32 times larger than in the sparsified case. This is clearly important to run efficient MD simulations. 
\subsection{Training procedure}
To train a MLP we optimize the variational parameters of the kernel $c_\mu$ with respect to a cost function, which is the sum of the Mean Squared Errors (MSE) of the total energies, the force components and the pressure plus a regularization term to prevent overfitting. Namely, for a given set of  configurations $R_1, \dots, R_N$, we have that the cost function $C$ is 
\begin{equation}
    \label{eq:cost-function}
C(c_\mu) = \alpha\mathrm{MSE}(E, \hat{E}) + \beta\mathrm{MSE}(F, \hat{F})+ \gamma\mathrm{MSE}(P, \hat{P}), 
\end{equation}
where $\alpha, \beta$ and $\gamma$ are adjustable prefactors, $E, F, P$ the reference energies, forces and pressures, $\hat{E}, \hat{F}, \hat{P}$ the predicted energies forces and pressures. Notice that in the above cost function for  the predicted quantities,  Eq.1 in the main text has to be evaluated for all local environments in the training set configurations, whereas the number of variational coefficients is much smaller as it  refers only to the sparsified local environments. But this remains a well defined least square fit, that has the advantage to avoid overfitting and/or working with a too large matrix.

In order to compute $\hat{F}$ and hence also $\hat{P}$ for a specific configuration, one has to compute the derivatives $\partial \hat{E}/\partial r_i$ of the total energy with respect to the atomic coordinates $r_i$. Here, we implement such derivatives with Adjoint Algorithmic Differentiation (AAD), which we also use to implement the derivatives of the cost function Eq.~\eqref{eq:cost-function} with respect to the variational parameters $c_\mu$. AAD, as opposed to finite differences methods, provides the exact evaluation of the derivatives at a cost which is within a factor of 4 the cost of evaluating the undifferentiated function. 

Since the cost $C$ is a quadratic function of the variational parameters $c_\mu$, taking the stationary equations leads to a linear system of the form 
\begin{align}\label{eq:linear-training}
    A_{\nu\mu}c_\mu = b_\nu,
\end{align}
where $\mu$ and $\nu$ range over the set of local environments, while the number of conditions to minimize the cost function $C(c_\mu)$ amounts to  $2N + 3\sum_iN_i$: $N$ conditions come from the first term of the cost function  in Eq.~\eqref{eq:cost-function}, from the comparison between predicted and true energies. Further $N$ conditions come from the third term of Eq.~\eqref{eq:cost-function}, which weighs the difference between predicted and true pressures. Finally the $3\sum_i N_i$ remaining conditions come from forces: for each configuration $i$ and each of the $N_i$ local environments we have 3 predicted force components, which we compare with the true ones.  Note that, $2N + 3\sum_iN_i$ is typically much larger than the number of variational parameters $c_\mu$ thanks to the adopted sparsification. Before solving the linear system (\ref{eq:linear-training}) a regularization for the matrix $A$ should be adopted in order to have not only a well conditioned inversion, but also to have a stable solution for the coefficients $c$. Here we have adopted the same one used for the energy optimization in VMC, \cite{2020NAK2}, namely the diagonal elements  of the matrix $A$ are scaled by $1+\lambda$, where $\lambda$ in this work has been set to $10^{-5}$. This type of regularization is different from the conventional one, where $A$ is substituted with  $A +I \lambda$. but is particularly important  here due to possible much different scales that may appear in the cost function for particular choices of $\alpha$,  $\beta$ and $\gamma$ (e.g. $\alpha \ll \gamma$). Moreover, since the dimension of the matrix $A_{\nu\mu}$ can become very large, we solve this system using the Conjugate Gradients Methods, 
via a parallel distributed implementation. 
We remark that  the present approach allows the solution of the  above linear system in a computational time that is only quadratic in the size of the training  data.

\subsection{Implementation details}
Our framework is a collection of \verb|Fortran| routines implementing: 
\begin{enumerate}
    \item the calculation of the discretized SOAP similarity measure between two local environments;
    \item the sparsification of a given dataset;
    \item the training of a a GKR model on a given (possibly sparsified) training set of configurations;
    \item the inference of a trained GKR model on a set of possibly new test configurations; this functionality is important to use the MLP obtained during training in a MD simulation.
\end{enumerate}
The code is parallelized using an efficient hybrid OpenMP/MPI paradigm. In the ideal resources allocation the number of MPI processes coincides with the number of configurations, whereas the OpenMP threading is used in the computation of the SOAP similarity measure, where the parallelization is over the number of atoms in a given configuration. Furthermore, we employ memory distribution in the training process: indeed, in that case, one has to solve the linear problem of Equation \eqref{eq:linear-training} and the matrix $A_{\mu\nu}$ can have considerable size even for relatively small datasets (\textit{e.g.} when using configurations containing a large number of atoms); each MPI process contains only selected rows on the matrix $A_{\mu\nu}$ and the linear problem is solved using Distributed Conjugate Gradients. Moreover, a regularization is applied so that the number of iterations required for the convergence of  Conjugate Gradients does not depend on the size of the matrix. 

At present our implementation can handle only configurations composed by one atomic species. Extending the code to multiple atomic species is part of ongoing research and, in the future, we plan to release our code as an open-source package.
\section{Validation tests}

\subsection{Model parameters and performances}
The kernel hyper-parameters used for the MLPs of this work are $\sigma = 0.5$ a.u. , $ n = \eta = 2$ (see notation in the main text). We have observed that a cutoff radius of $r_c = 3.5$ a.u. is appropriate for this system. To test this we have trained a model using a larger cutoff ($r_c = 5.0$ a.u) and verified that the two models give the same EOS. 
During the training procedure we have used different values of the prefactors in Eq.~\eqref{eq:cost-function}. In particular, when we trained the MLPs using all the contributions, we used $\alpha = 0.1$, $\beta = 1$ and $\gamma = 100$ in order to take into account the different scales of energy, forces and pressure. For the training with energy only we used $\alpha = 1$, $\beta = \gamma = 0$. The whole dataset was divided into training and test set, the latter one being extracted from the former using stratified sampling with respect to $r_{s}$. The final model used for the MD simulations has the following performances on the {\it test} set: 1. RMSE on energies $2.197\times 10^{-2}$; 2. RMSE of forces $2.451\times 10^{-3}$; 3. RMSE on pressures  $1.766\times 10^{-6}$.

\subsection{Training only with energies}
\begin{figure}[th]
\includegraphics[scale=0.5]{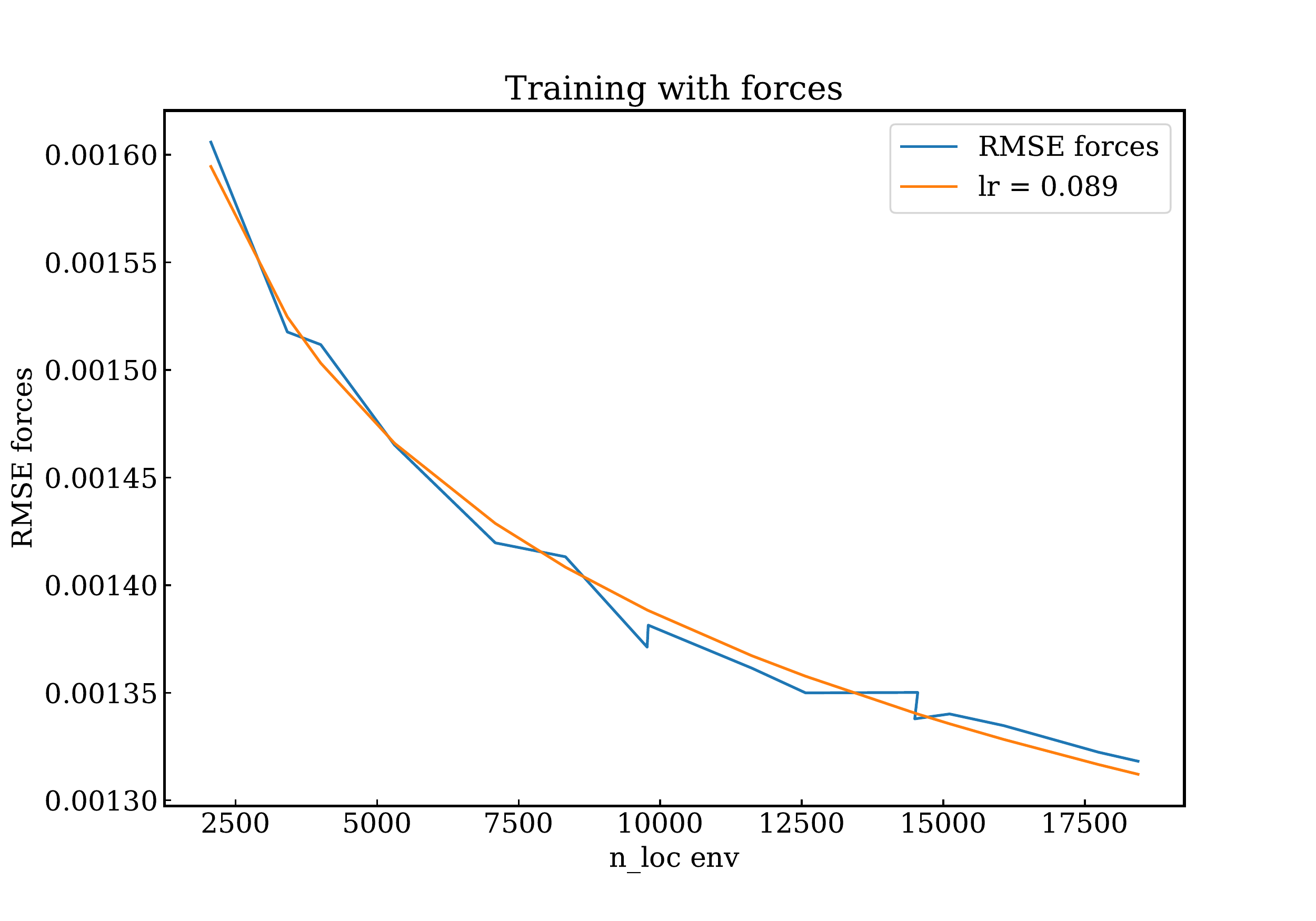}
\includegraphics[scale=0.5]{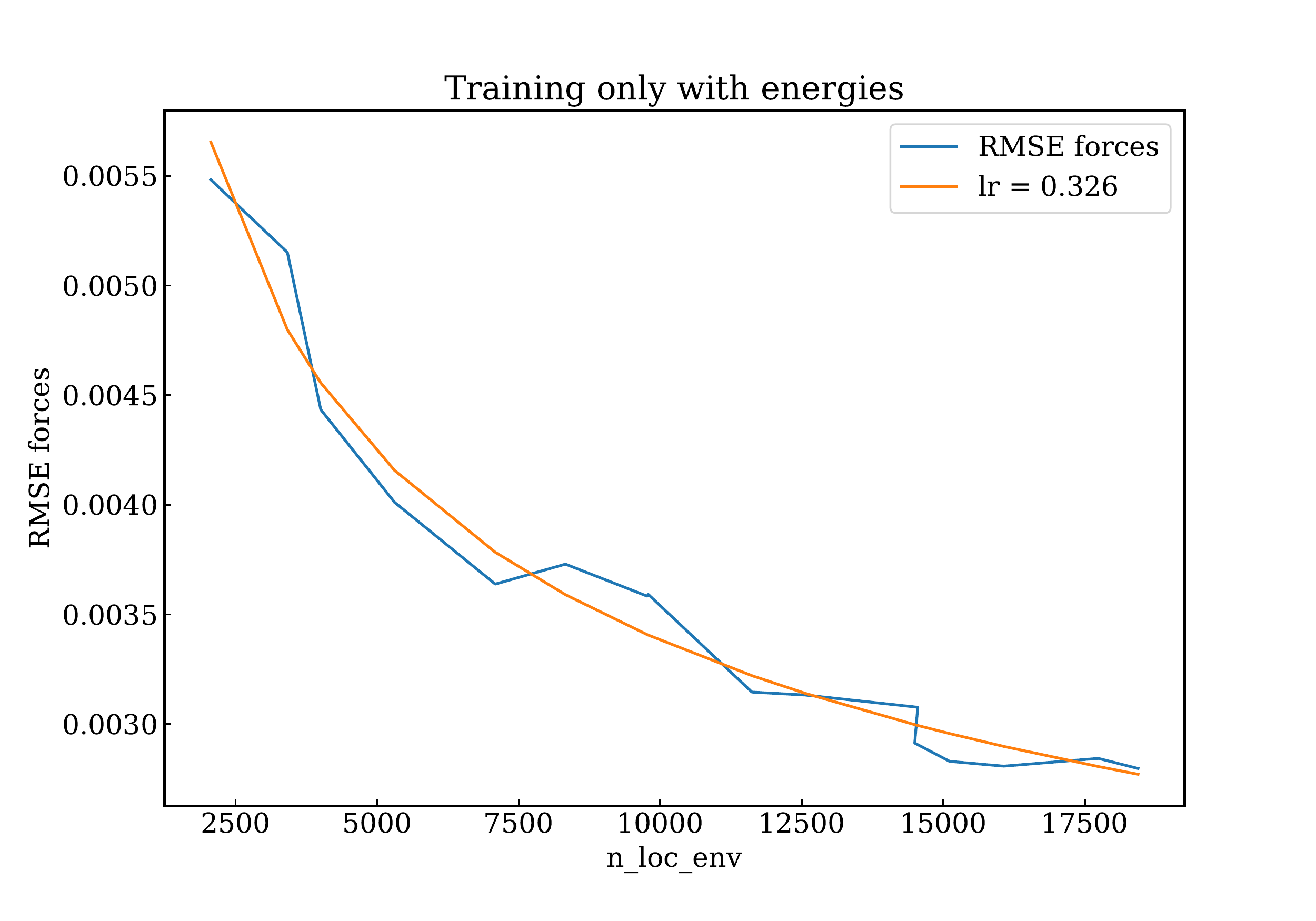}
\caption{Comparing the learning rates for the RMSE on forces for models MLP$_1$ and MLP$_2$. Training with forces, upper panel; training only with energies, lower panel. The reported RMSEs are evaluated on the test set, none of whose configurations has been seen by the model during training. }
\label{fig:only_energies}
\end{figure}

An important validation test to perform is to compare the difference in performances when training a MLP with or without forces and pressures. Indeed, training only with energies has the advantage of being less computationally demanding; this often comes at a price: the performances of a model obtained in this way, especially when looking at the RMSE on forces and pressure, might not be satisfactory. To get a clear evidence of this we first trained a MLP, referred to as MLP$_1$, using only energies and then added forces components and pressures, obtaining a new model, name MLP$_2$.  Both models are trained on the difference between VMC and PBE training data. We compared the learning curves for MLP$_1$ and MLP$_2$. While MLP$_1$ displays a much higher learning rate for all the assessed metrics, its performances are significantly worse than those of MLP$_2$. One can also easily estimate that, in order for MLP$_1$ to reach the same performances of MLP$_2$, one would need a much larger dataset, with a number of local environments more than an order of magnitude larger than the ones present in the dataset used for the validation tests. Even if this could be possible (which, for VMC, would be extremely costly), this would come with efficiency issues: indeed, the inference time for a GKR models scales linearly with the size of the dataset; thus, a MLP trained on such a large dataset would be unfeasible to use in MD simulations. This confirms the importance of using forces and pressures during training. These considerations are supported by Fig. \ref{fig:only_energies}, where we compare the learning rates of MLP$_1$ and MLP$_2$ for the RMSE on forces. By performing a straightforward interpolation, we see that for MLP$_1$ to have the same performances of MLP$_2$ almost 200 thousands local environments in the training set would be necessary. 

\subsection{Effect of noisy data on model performances}
One significant difference between our work and previous ones is the fact that we trained a MLP using a stochastic \textit{ab-initio method}, while most of the papers in the literature so far have focused on deterministic methods such as DFT. It is therefore important to understand how the stochasticity of a method like QMC affects the performances of a MLP: To this end, we trained two types of MLPs, one on the difference BLYP $-$ PBE, which we call MLP$_1$, and the other on the difference VMC $-$ PBE, called MLP$_2$. To better appreciate the performances of the two models, we list in Table \ref{tab:initial-rmses} the initial RMSEs for the two collections of training sets.
\begin{table}[ht!]
  \centering
\begin{tabular}{||r|r|r|r||}
\hline
 Dataset &  RMSE$_e$ &  RMSE$_p$ &  RMSE$_f$\\
\hline
     VMC - PBE &   2.94E-01 & 2.49E-04 & 4.75E-03\\
     BLYP - PBE &  1.2E-00 &  4.21E-04 & 5.45E-03 \\
\hline
\end{tabular}
  \cprotect\caption{Initial RSMEs on VMC - PBE and BLYP - PBE. The reported RMSEs are evaluated on the test set, none of whose configurations has been seen by the model during training. }
  \label{tab:initial-rmses}
\end{table}

\begin{figure}[th]
\includegraphics[width=14cm]{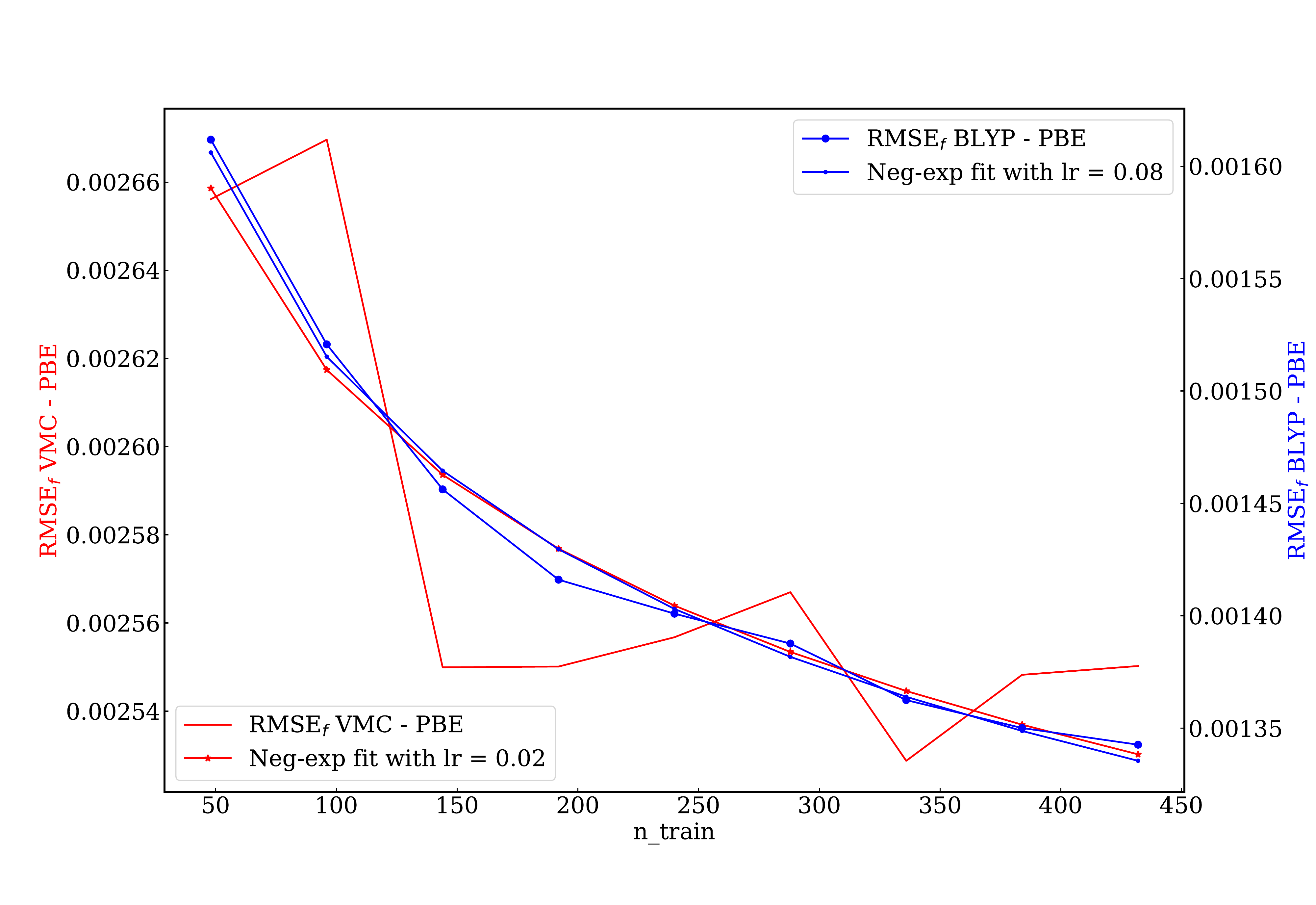}
\caption{Comparison of learning rates on force prediction for MLP$_1$ and MLP$_2$;  all the RMSEs are evaluated on the test set, none of whose configurations has been seen by the model during training. }
\label{fig:learning_rates} 
\end{figure}

In Table \ref{tab:noise-effect}, we show the comparison between the performances of MLP$_1$ and MLP$_2$: we notice that MLP$_1$ performs significantly better in predicting all the observables. Most importantly, MLP$_1$ can be (for the largest training set) twice as accurate than MLP$_2$ in predicting forces. Moreover, in Fig. \ref{fig:learning_rates} we show the comparison between the learning rate curves for force prediction of MLP$_1$ and MLP$_2$, respectively: the learning rate for forces of MLP$_1$ is 4 times bigger than the corresponding value for MLP$_2$.  Thus, we can conclude that training a MLP on stochastic data, such as the one generated in QMC, is a much more difficult task, both in terms of model performances and amount of data required to reach a desired accuracy. 

\begin{table}[ht!]
  \centering
\begin{tabular}{||r|r|r|r|r|r|r|r||}
\hline
 ntrain &  ntest &  RMSE$^{{\rm{MLP}}_1}_e$ &  RMSE$^{{\rm{MLP}}_1}_p$ &  RMSE$^{{\rm{MLP}}_1}_f$ &  RMSE$^{{\rm{MLP}}_2}_e$ &  RMSE$^{{\rm{MLP}}_2}_p$ &  RMSE$^{{\rm{MLP}}_2}_f$ \\
\hline
     48 &     30 &         1.10E-02 &         2.59E-07 &         1.61E-03 &         9.75E-02 &         1.81E-06 &         2.66E-03 \\
     96 &     30 &         1.06E-02 &         2.22E-07 &         1.52E-03 &         7.44E-02 &         2.43E-06 &         2.67E-03 \\
    144 &     30 &         1.29E-02 &         1.29E-07 &         1.46E-03 &         8.76E-02 &         5.78E-07 &         2.55E-03 \\
    192 &     30 &         1.18E-02 &         1.76E-07 &         1.42E-03 &         7.62E-02 &         7.26E-07 &         2.55E-03 \\
    240 &     30 &         1.13E-02 &         1.44E-07 &         1.40E-03 &         7.52E-02 &         8.63E-07 &         2.56E-03 \\
    288 &     30 &         1.08E-02 &         1.48E-07 &         1.39E-03 &         6.73E-02 &         9.79E-07 &         2.57E-03 \\
    336 &     30 &         1.16E-02 &         1.46E-07 &         1.36E-03 &         7.05E-02 &         9.77E-07 &         2.53E-03 \\
    384 &     30 &         1.20E-02 &         1.34E-07 &         1.35E-03 &         6.61E-02 &         1.21E-06 &         2.55E-03 \\
    432 &     30 &         1.19E-02 &         1.39E-07 &         1.34E-03 &         7.02E-02 &         8.25E-07 &         2.55E-03 \\
\hline
\end{tabular}
  \cprotect\caption{Comparison between BLYP - PBE (MLP$_1$) and VMC - PBE (MLP$_2$); all the RMSEs are evaluated on the test set, none of whose configurations has been seen by the model during training. }
  \label{tab:noise-effect}
\end{table}

\subsection{Transferability}
\begin{figure}[th]
\includegraphics[width=14cm]{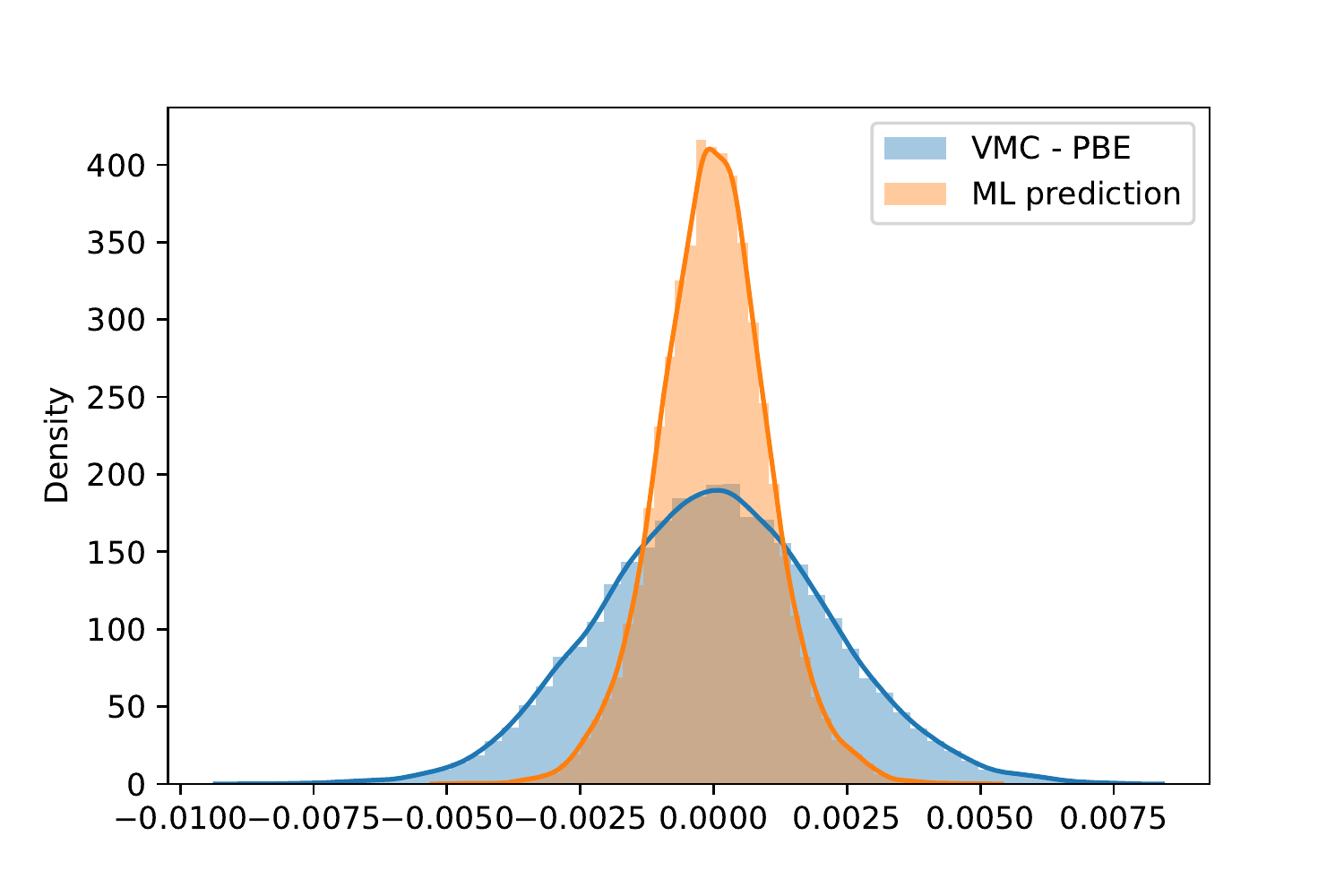}
\caption{Distributions of the difference force components: the blue one represents the ``true" difference between VMC and PBE force components, while the orange distribution is the ML correction.
Indeed, $x$ axis refers to the difference between VMC and PBE force components, $\Delta F_{i} = F_{i}^{\rm{VMC}} - F_{i}^{\rm{PBE}}$ for ``VMC-PBE", while $x$ axis refers to the discrepancy between a ML prediction and the corresponding target value, $\Delta F_{i} = \left( \Delta F_{i}^{\rm{ML-correction}} - (F_{i}^{\rm{VMC}} - F_{i}^{\rm{PBE}}) \right)$ for ``ML prediction". Here, the former result shows the fact that VMC and PBE forces are distinctively different, while the latter one indicates that our ML model corrects the PBE forces in the right direction (i.e, to the VMC forces).}
\label{fig:diff_256}
\end{figure}

Another important aspect to consider in evaluating a MLP is transferability, \textit{i.e.} the capability of the model to make accurate predictions on configurations whose size has not been encountered in the training set.  In our case, we tested a MLP trained on a dataset composed by configurations with 128 hydrogen atoms on a test set made of 25 configurations of 256 hydrogen atoms. The MLP was trained on the difference between VMC and PBE energies, forces and pressures. On the test set, the RMSE between PBE and VMC forces is $2.1\times 10^{-3}$ Ha/bohr. Even on such a different dataset the MLP performs very well, with an RMSE on forces of $1.6\times 10^{-3}$ Ha/bohr. Moreover, in Fig. \ref{fig:diff_256}, plotting the distribution of the difference between VMC and PBE force components and the distribution of the MLP predictions, we see a distinctive difference, showing that the model significantly corrects the PBE initial simulation.

\begin{figure*}[th]
    \centering
    \begin{subfigure}{.48\textwidth}
  \centering
  \includegraphics[width=\linewidth]{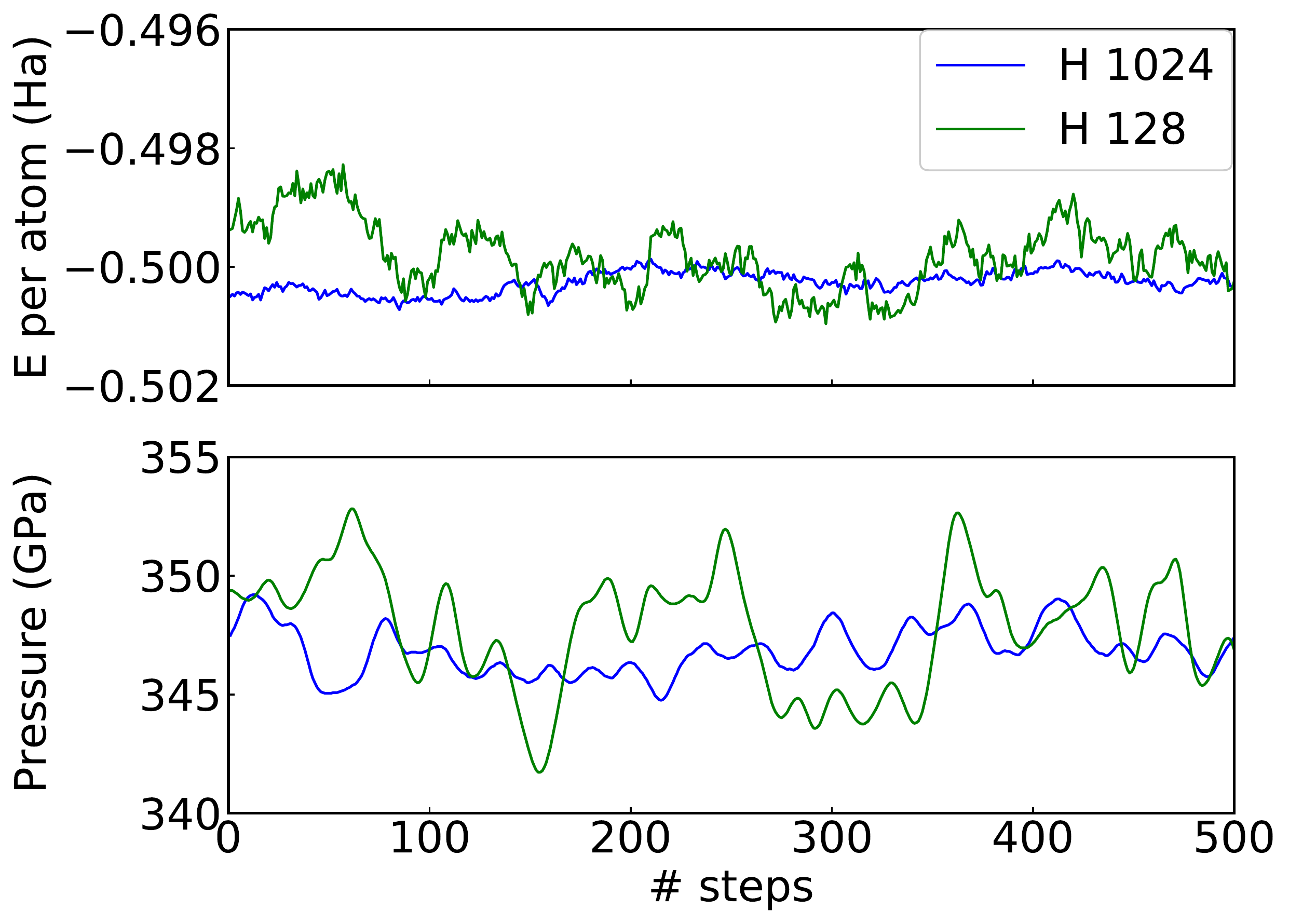}
  \caption{}
  \label{fig:1024_rs1.30}
\end{subfigure}
    \begin{subfigure}{.48\textwidth}
  \centering
  \includegraphics[width=\linewidth]{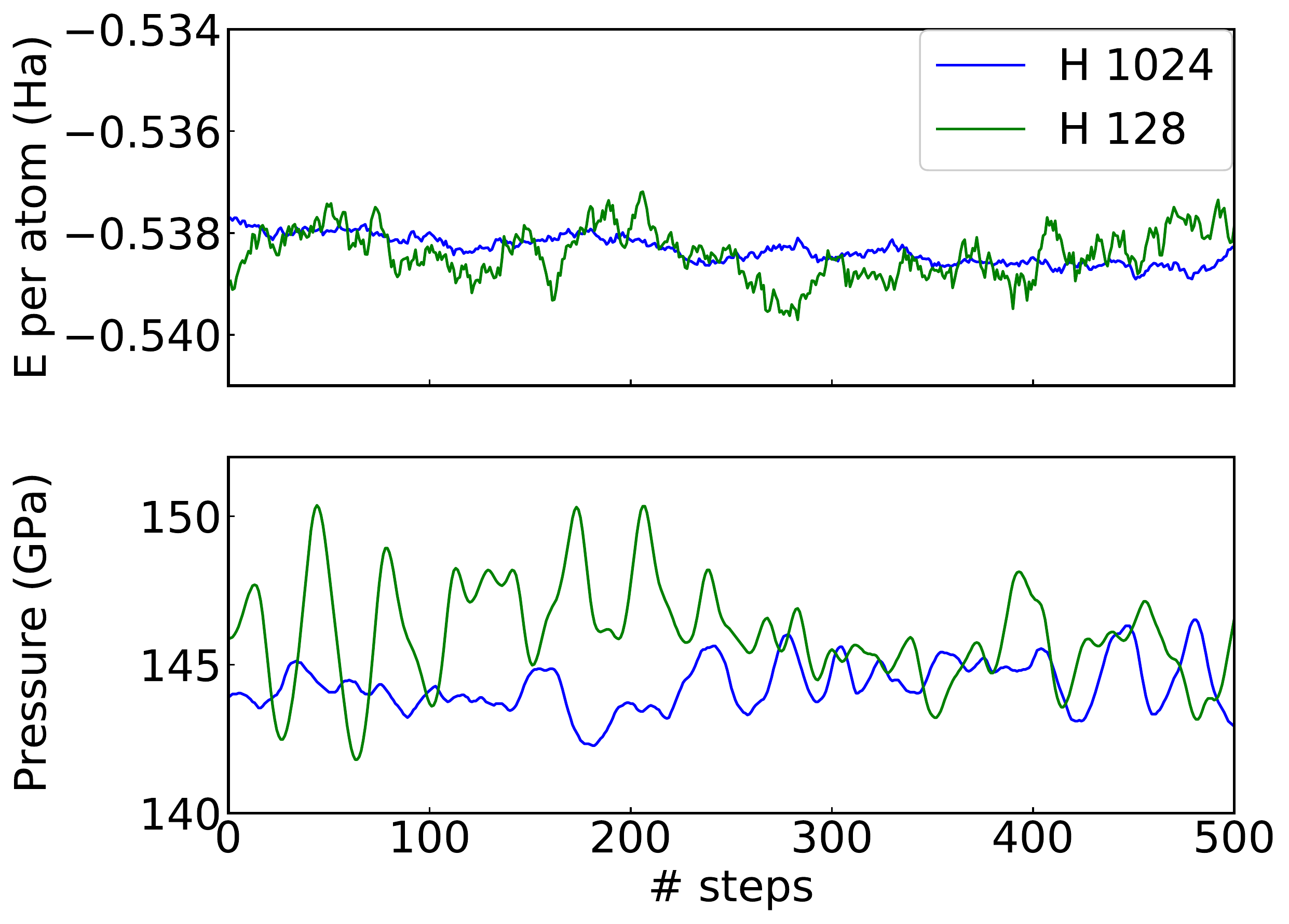} 
  \caption{}
  \label{fig:1024_rs1.50}
\end{subfigure}
\caption{Instantaneous values of the energy per atom (upper panel) and pressure (lower panel) during MD simulations at 1200~K for H 128 (green lines) and H 1024 atoms (blue lines). We employed two WS radii, (a) $r_{\rm{s}} = 1.30$ a.u and (b) $r_{\rm{s}} = 1.50$ a.u.}
\label{fig:1024_comparison}
\end{figure*}

Given the transferability of our MLP, we performed MD simulations on larger systems to study possible finite size effects. Here we report the results of two simulations of a system with 1024 atoms at 1200~K and with Wigner-Seitz radii of $r_{\rm{s}} = 1.30$ a.u. and $r_{\rm{s}} = 1.50$ a.u. respectively. Notice that these simulations are computationally prohibitive with the standard VMC MD approach. As shown in Fig.~\ref{fig:1024_comparison} thermodynamic quantities such as the pressure and the total energy per atom seem to be weakly dependent on the size of the system, at least for densities far enough from the transition.

\bibliography{references}